\documentclass[prl,aps,floats,superscriptaddress,twocolumn]{revtex4-1}

\usepackage{amsmath}
\usepackage{amssymb}
\usepackage{latexsym}
\usepackage[english]{babel}
\usepackage{blindtext}
\usepackage{graphicx}
\usepackage{subfigure}
\usepackage{indentfirst}
\usepackage{appendix}
\usepackage{mathrsfs}
\usepackage{color}
\usepackage[colorlinks,linkcolor=blue,anchorcolor=blue,citecolor=green]{hyperref}
\usepackage{ulem}
\usepackage{float}
\usepackage{subfiles}

\allowdisplaybreaks[2]

\graphicspath{{figure/}}

\newcommand {\no} {\nonumber}

\newcommand{\lt} {\left}
\newcommand{\rt} {\right}

\makeatother

\begin{document}

\title{Characterization of collective excitations in weakly-coupled disordered superconductors}
\author{Bo Fan}
\email{bo.fan@sjtu.edu.cn}
\affiliation{Shanghai Center for Complex Physics, 
	School of Physics and Astronomy, Shanghai Jiao Tong
	University, Shanghai 200240, China}
\author{Abhisek Samanta}\email{abhiseks@campus.technion.ac.il}
\affiliation{ Physics Department, Technion, Haifa 32000, Israel}
\author{Antonio M. Garc\'ia-Garc\'ia}
\email{amgg@sjtu.edu.cn}
\affiliation{Shanghai Center for Complex Physics, 
	School of Physics and Astronomy, Shanghai Jiao Tong
	University, Shanghai 200240, China}

\date{\today }

\begin{abstract}
	Isolated islands in two-dimensional strongly-disordered and strongly-coupled superconductors become optically active inducing sub-gap collective excitations in the ac conductivity. Here, we investigate the fate of these excitations as a function of the disorder strength in the  experimentally relevant case of weak electron-phonon coupling. An explicit calculation of the ac conductivity, that includes vertex corrections to restore gauge symmetry, reveals the existence of collective sub-gap excitations, related to phase fluctuations and therefore identified as the Goldstone modes, for intermediate to strong disorder. As disorder increases, the shape of the sub-gap excitation transits from peaked close to the spectral gap to a broader distribution reaching much smaller frequencies. Phase-coherence still holds in part of this disorder regime. The requirement to observe sub-gap excitations is not the existence of isolated islands acting as nano-antennas but rather the combination of a sufficiently inhomogeneous order parameter with a phase fluctuation correlation length smaller than the system size. Our results indicate that, by tuning disorder, the Goldstone mode may be observed experimentally in metallic superconductors based for instance on Al, Sn, Pb or Nb.  
\end{abstract}
\maketitle
\newpage

P. W. Anderson stated  \cite{cooper2010} that Bardeen-Cooper-Schrieffer (BCS) theory of superconductivity \cite{Bardeen1957} had been the scientific love of his life. It is likely that collective modes were an important part of this love story. Shortly after the microscopic BCS theory \cite{Bardeen1957} was proposed, Anderson \cite{anderson1958sc,anderson1963} noticed that two of its most salient features, the existence of a gapped ground state and phase rigidity were to some extent contradictory. If the phase of the order parameter were rigid, the $U(1)$ gauge symmetry is spontaneously broken.
According to Goldstone's theorem \cite{nambu1960,goldstone1961}, the spontaneously breaking of this $U(1)$ symmetry 
is associated to the existence of a zero energy (massless) collective excitation, the so called Goldstone mode. In principle, this is in tension with the BCS prediction of a gapped ground state.  
However, Anderson argued \cite{anderson1958sc} that for clean superconductors, later \cite{belitz1989} shown to also hold for weakly disordered superconductors,  the Goldstone mode is not observable because long range Coulomb interactions shift its natural frequency to the plasmon frequency which is typically much higher than the spectroscopic gap. 

Therefore, it came as a relative surprise that recent numerical results for the conductivity of two-dimensional strongly-disordered and strongly-coupled superconductors,\cite{cea2014,cea2015,Seibold2015,barabash2003,sherman2015,swanson2014,seibold2017,abhisek2020} have shown the existence of collective excitations below the spectral gap. The optical absorption of the incoming electromagnetic radiation occurs \cite{cea2014} in disorder-induced isolated superconducting islands that act like nano-antennas. 
The combination of strong disorder and strong coupling mixes zero and finite momentum modes so that collective modes contribute to the optical response even in the long wavelength limit. 
Moreover, it was argued \cite{cea2014,belitz1989} that long range Coulomb interactions do not change this conclusion qualitatively. 
Although these \cite{cea2014} numerical results provide rather conclusive evidence on the existence of sub-gap collective excitations, they were obtained in the strong-coupling limit which is not strictly applicable in most metallic superconductors such as Sn, Nb, Al or Pb whose electron-phonon coupling is weak or intermediate.  

On the experimental front, there are also recent observations of sub-gap structure in the optical conductivity of several disordered weakly-coupled superconductors \cite{crane2007,Driessen2012,sherman2015,neilinger2015,pracht2016}, see also Refs.~\cite{Mondal2011,Chand2012,Mondal2013,cheng2016,Orr1985,Jaeger1989,goldman1993,thiemann2018,graybeal1984,shahar1992} for related developments. In NbN and InO \cite{sherman2015,matsunaga2014} close to the superconductor-insulator transition, sub-gap weight has been related to amplitude fluctuations, the Higgs mode \cite{shimano2020}. In granular aluminum \cite{pracht2017}, the observation of spectral weight below the gap at relatively high temperature was associated with the Goldstone mode though the agreement with the theoretical predictions was only qualitative. Another experiment \cite{levy2019} involving granular aluminum, performed at lower temperatures, reported a broad sub-gap peak whose origin remains unexplained. The conclusion is that, despite promising advances, there is no yet conclusive evidence that the different sub-gap excitations observed experimentally are the sought Goldstone and Higgs modes due to both the qualitative nature of the theoretical predictions and the difficulty in ruling out other experimental causes such as the effect of the substrate or competing quantum orders.
\begin{figure}
		\includegraphics[width=5.cm]{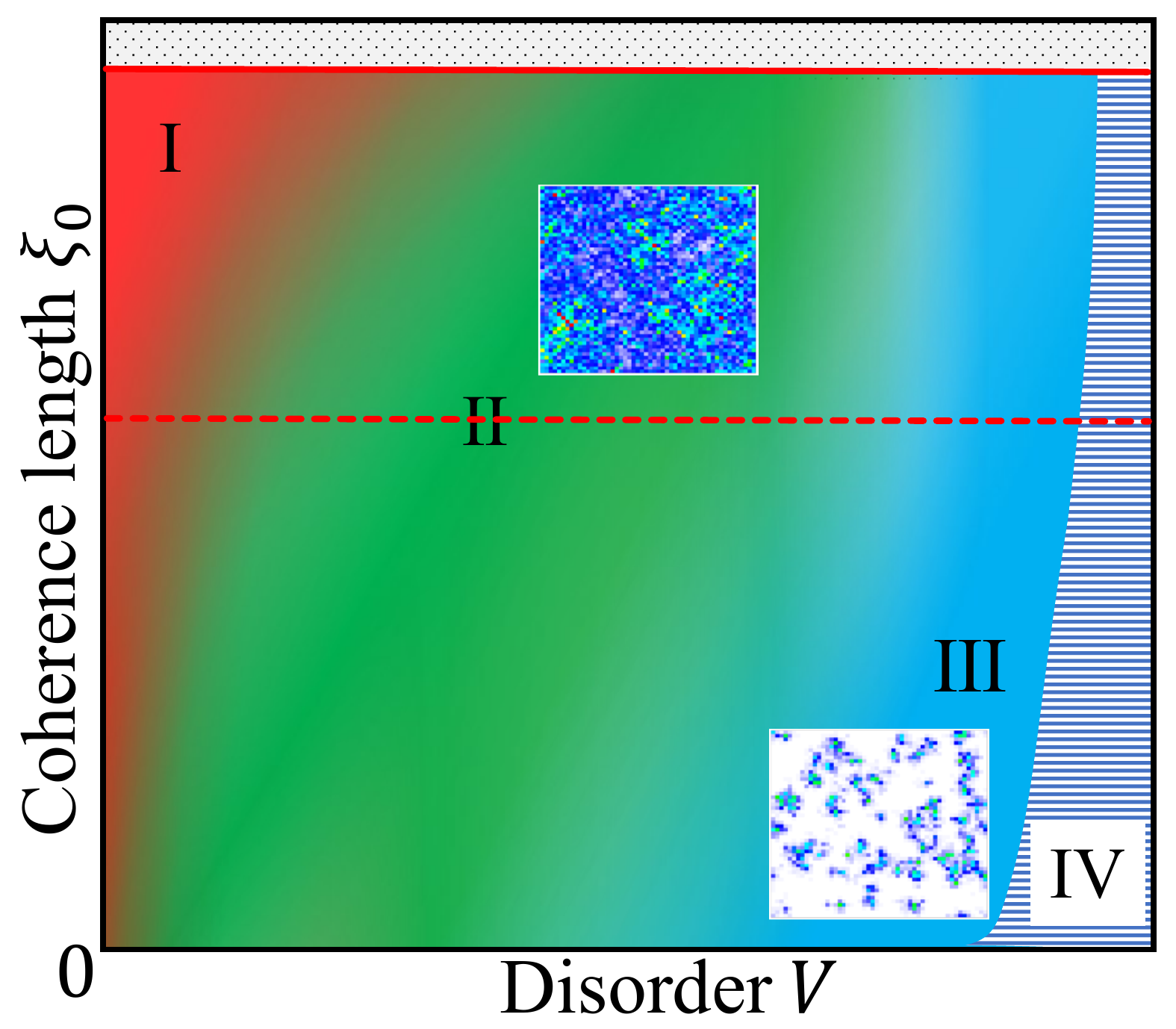}
		\caption{Summary of the sub-gap optical conductivity as a function of disorder $V$ and the electron-phonon coupling strength parameterized by the clean coherence length $\xi_0$. 
			\uppercase\expandafter{\romannumeral1} (red): Region of very weak disorder where no collective excitations are expected either due to Coulomb interactions or because the collective mode is still gapless. \uppercase\expandafter{\romannumeral2} (green): Collective excitations can be observed and only require a sufficiently inhomogeneous order parameter. We illustrate it with a mini-plot of the site dependence of the order parameter for $U=1, V=1.5$. The dashed red line separates the weak coupling (above) and strong coupling (below) regions. Numerically, we explore the range $\xi_0 \lesssim 500$nm that includes most weakly-coupled metallic superconductors (above the dashed line).  
			\uppercase\expandafter{\romannumeral3} (blue): Collective excitations are related to isolated superconducting islands. Here, the mini plot is for $U=5, V=3$. The strong coupling limit was previously studied in  Ref.~\cite{cea2014}.  \uppercase\expandafter{\romannumeral4}: Anderson insulator region. The top grey dotted region $\xi_0 \rightarrow \infty$ is not accessible numerically.} \label{Fig.phase_diagram}
\end{figure}

In this paper, we investigate collective excitations in a fermionic model of two-dimensional disordered superconductors focusing on the optical response captured by the low-frequency ac conductivity. Our analysis is based on the Bogoliubov-de Gennes mean-field formalism which leads to the so-called {\it bare bubble} diagram in the calculation of the conductivity, plus its vertex corrections \cite{schrieffer2018} which includes fluctuations around the mean-field order parameter (namely amplitude, phase, and density fluctuations) evaluated within the random phase approximation \cite{anderson1958sc,Supplementary}.
This is the minimal calculation scheme that restores gauge invariance and therefore can describe collective excitations. We reach larger system sizes which allow us to explore the weak coupling limit. 

In Fig.~\ref{Fig.phase_diagram}, we sketch the pattern of sub-gap excitations in the ac conductivity as a function of the strength of disorder and electron-phonon coupling. The main results of the paper corresponds to region II (green), especially above the dashed red line, where we identify the Goldstone mode in weakly coupled superconductors whose existence only requires a sufficiently inhomogeneous  \cite{ma1985,ghosal1998,Ghosal2001,mayoh2015global,bofan2020,bofan2020a,verdu2018,Burmistrov2012,Gastiasoro2018} order parameter. Region III (blue), corresponds to the region where sub-gap optical response is related to isolated islands \cite{cea2014}. The strong coupling region was previously studied in Ref.~\cite{cea2014}. \\
\begin{figure}
		\subfigure[]{\label{fig.sptial_U1_1}
			\includegraphics[width=2.6cm]{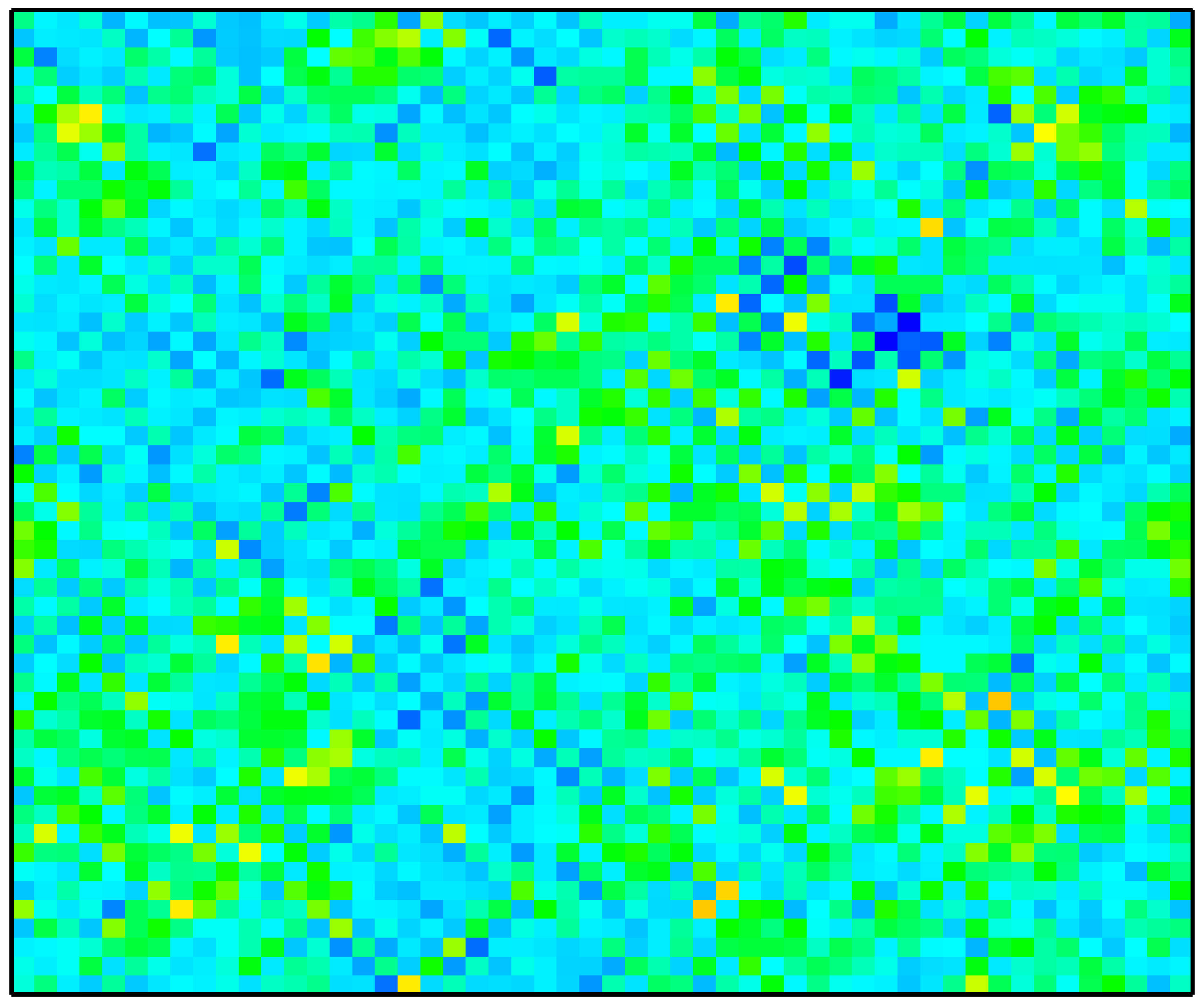}}
		\subfigure[]{\label{fig.sptial_U1_2}
			\includegraphics[width=2.6cm]{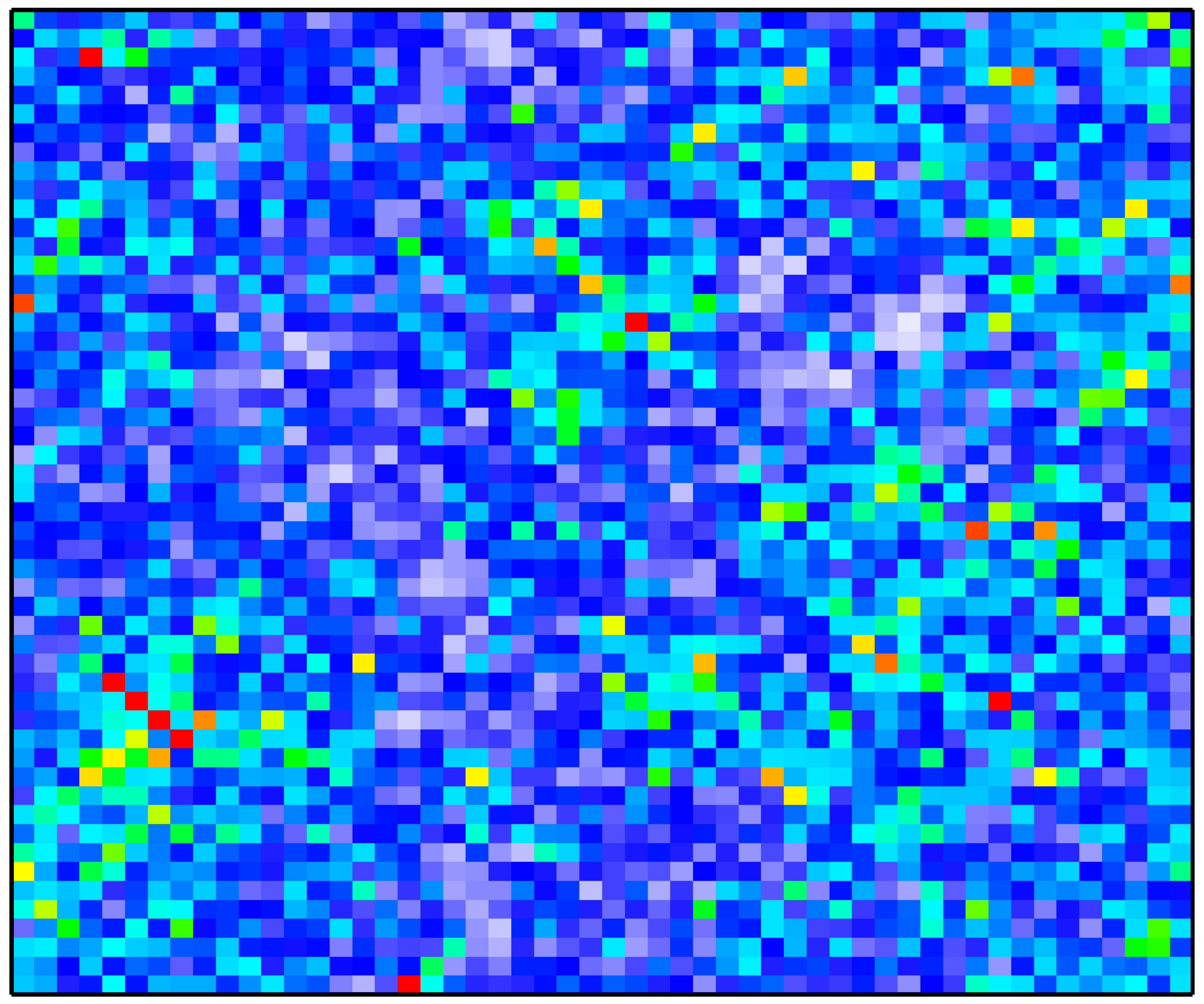}}
		\subfigure[]{\label{fig.sptial_U1_3}
			\includegraphics[width=2.6cm]{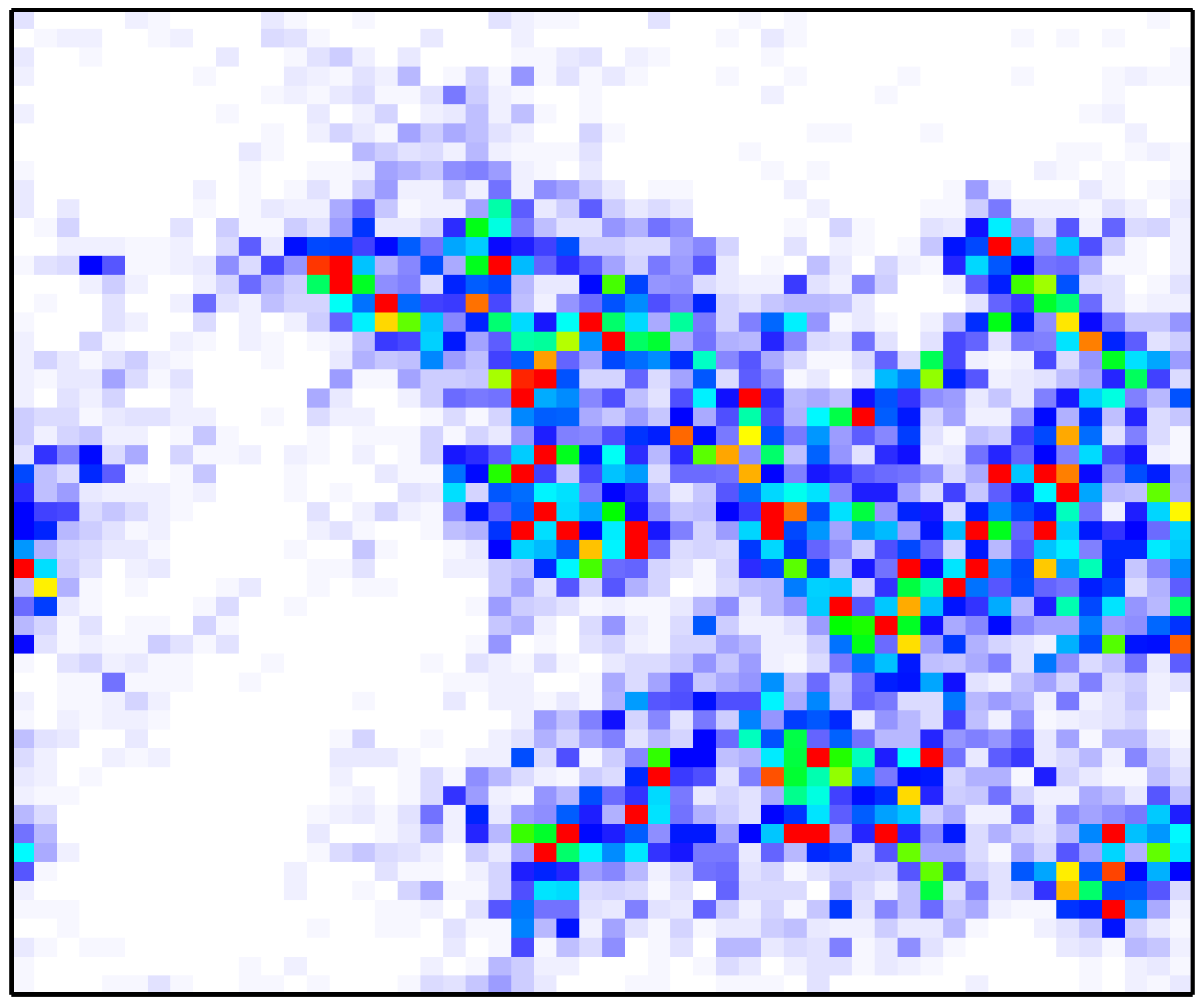}}\\
		\subfigure[]{\label{fig.sptial_U5_1}
			\includegraphics[width=2.6cm]{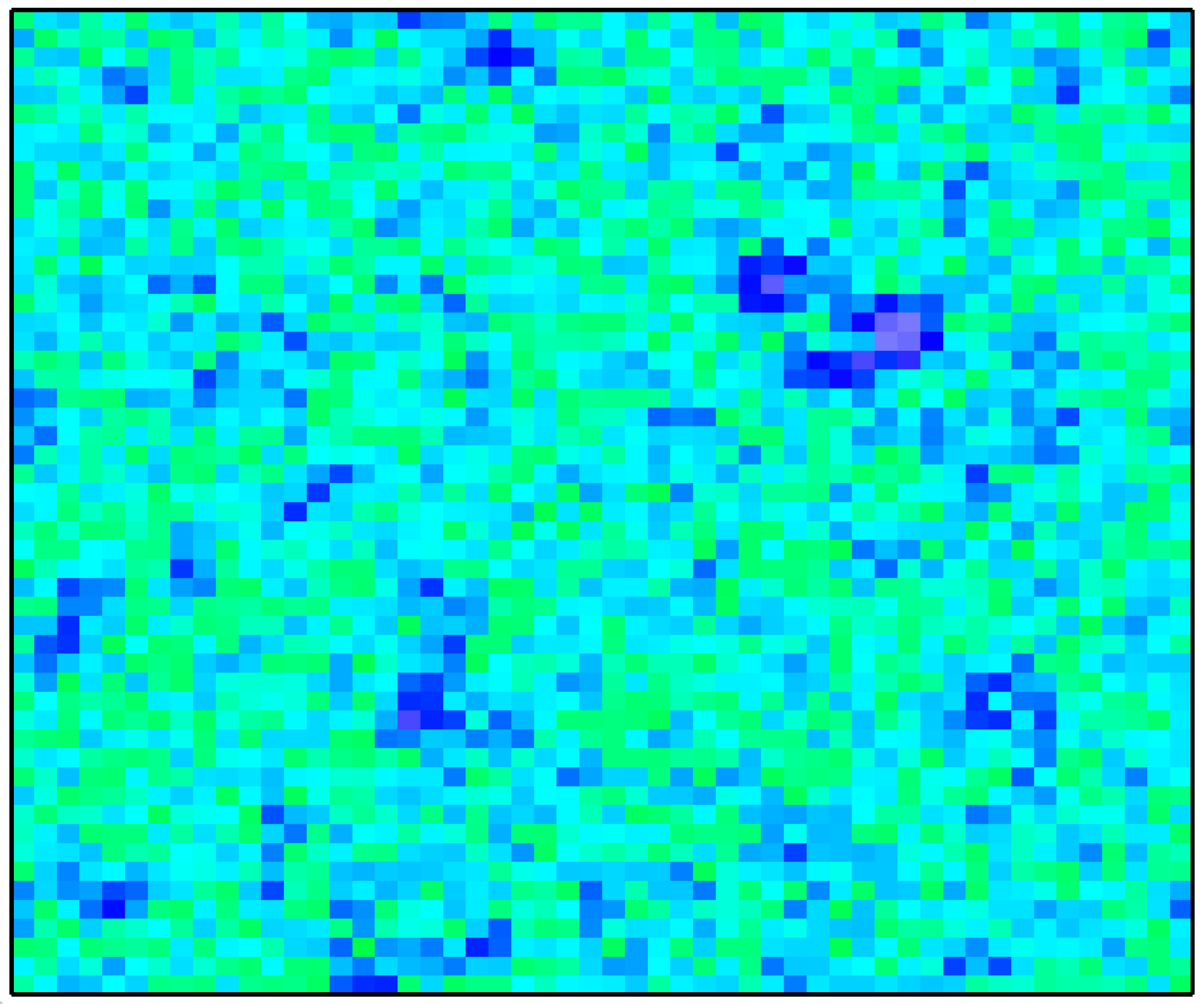}}
		\subfigure[]{\label{fig.sptial_U5_2}
			\includegraphics[width=2.6cm]{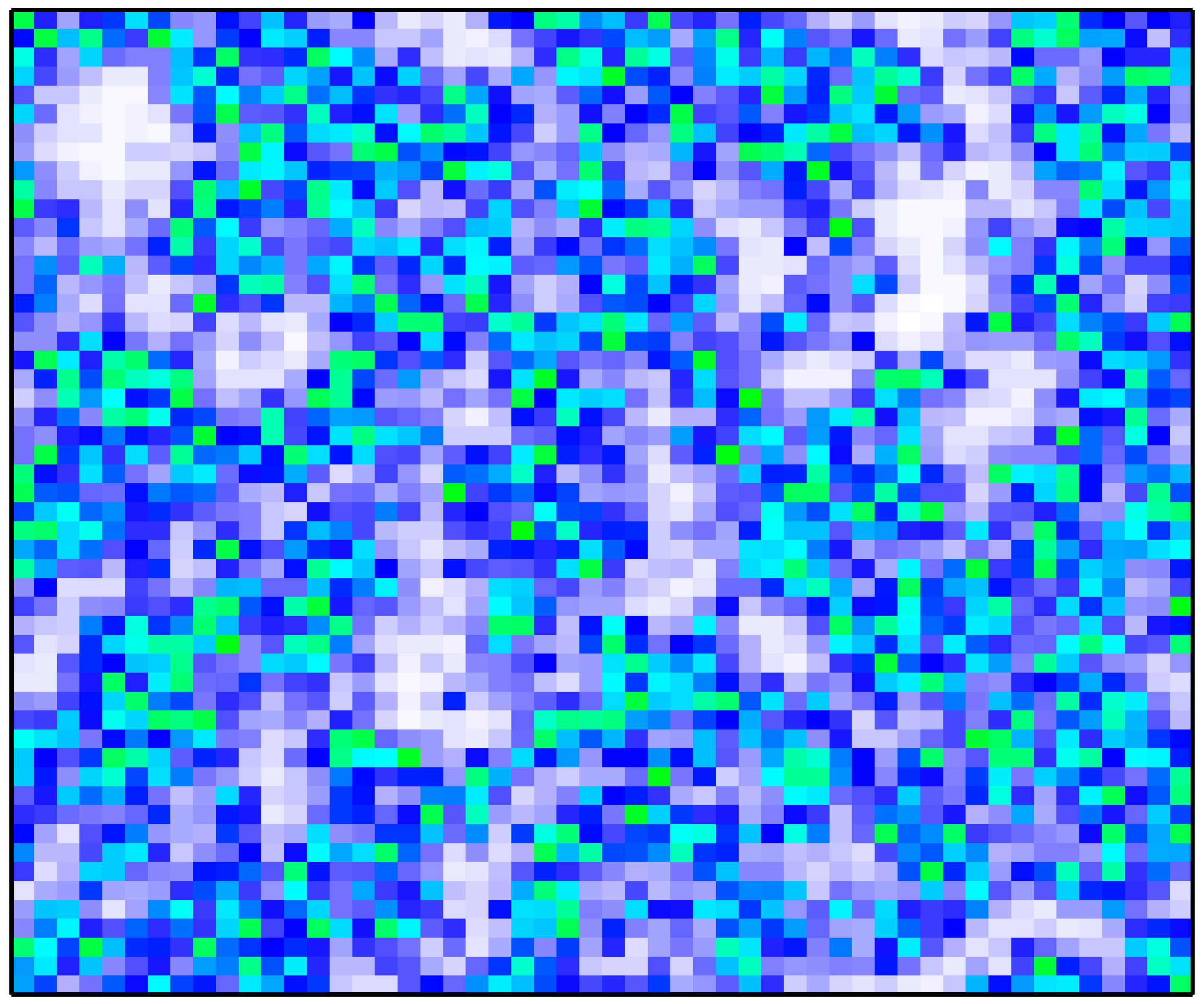}}
		\subfigure[]{\label{fig.sptial_U5_3}
			\includegraphics[width=2.6cm]{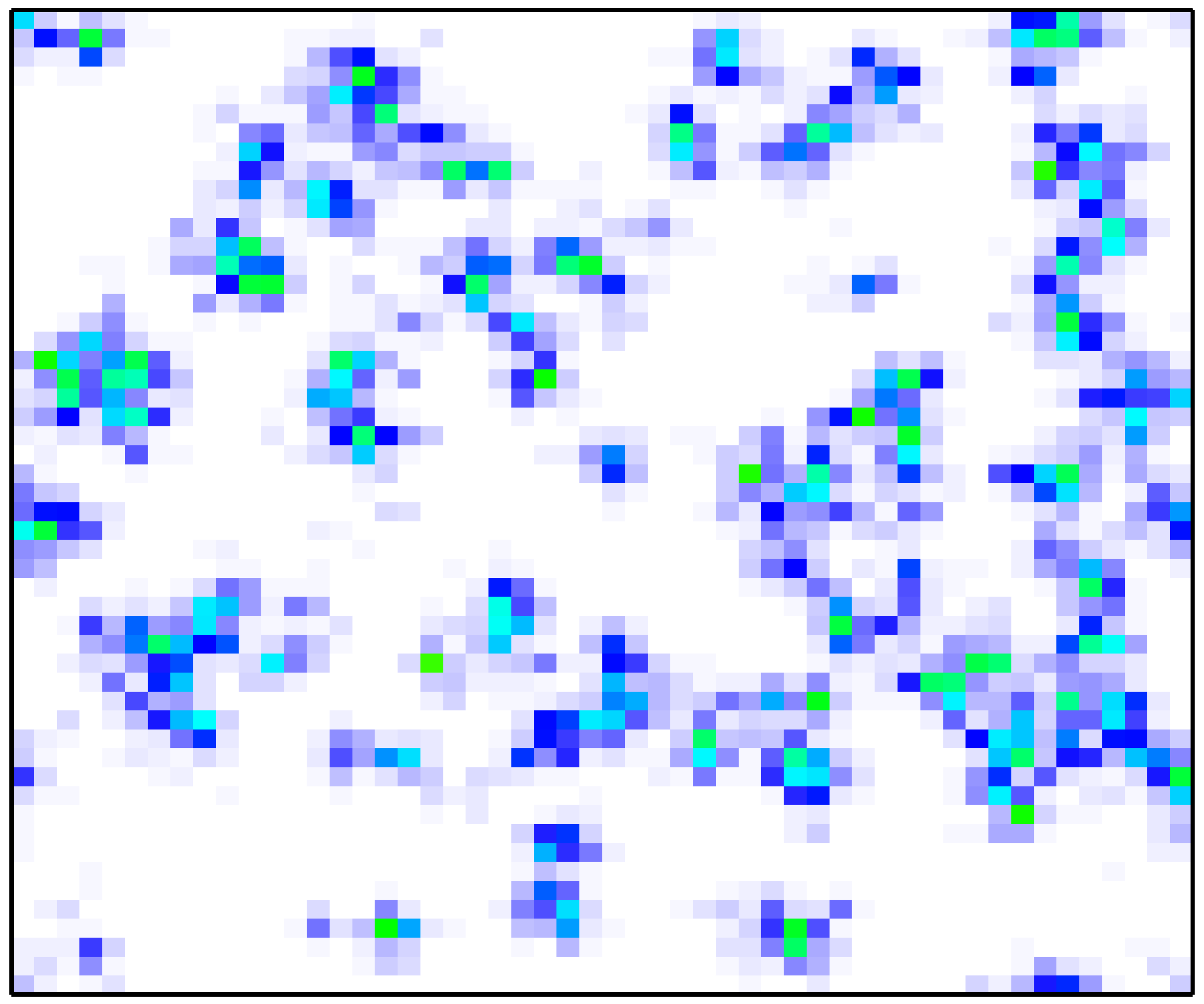}}\\
		\includegraphics[width=3cm]{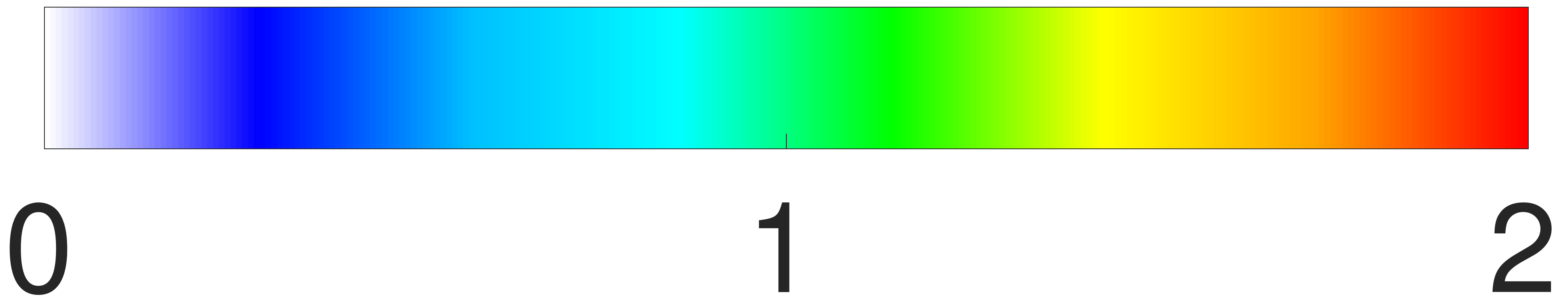}
		\caption{Amplitude of the order parameter $\Delta(r_i)$ resulting from the solution of the BdG equations, normalized by its value $\Delta_0$ in the clean limit. The spatial maps are plotted on a square lattice of size $N=52\times52$, with $\langle n \rangle =0.875$. Upper: $U=1$. Lower: $U=5$. Disorder strength, from left to right, is $V = 0.5, 1.5$ and $3$.} \label{Fig.spatial}
\end{figure}

We initiate our analysis with a brief summary of the employed theoretical framework leading to the calculation of the ac conductivity (see supplemental information \cite{Supplementary} for details). The first step is the numerical calculation of the eigenvalues $E_n$ and eigenvectors $[u_n(r_i), v_n(r_i)]$  
of the two-dimensional Bogoliubov de-Gennes (BdG) equations in the presence of a random potential \cite{DeGennes1964, DeGennes1966, Ghosal2001, ghosal1998}
with an uniform distribution $[-V,V]$, an electron-phonon coupling $U$ and a chemical potential  $\mu_i = \mu + Un(r_i)/2$ that incorporates a site-dependent Hartree shift. The BdG equations are 
completed by self-consistency conditions for the site-dependent order parameter amplitude $\Delta(r_i) = U\sum_{n}u_n(r_i)v_n^*(r_i)$ and the density $n(r_i) = 2\sum_{n}|v_n(r_i)|^2$,
that are also outputs of the numerical calculation. 
All the presented results are for a square lattice $(N=L\times L)$ with periodic or Dirichlet boundary conditions. We fix the averaged density $\langle n \rangle =\sum_{r}n(r)/N$ and let the chemical potential $\mu$ vary.
As an example of the BdG solution, that illustrates the differences between weak $U =1$ and strong $U=5$ coupling, in Fig.~\ref{Fig.spatial}, we depict $\Delta(r_i)$ in these two regimes. For strong disorder, the order parameter is distributed in small islands, while for weak coupling we observe an intricate, highly inhomogeneous spatial structure with no visible islands. This stark difference will be important in the following analysis of the conductivity. 

\indent The second step of the calculation is the evaluation of the response function \cite{cea2014},
$\chi_{ij}(j^x,j^x)=-i \int dt e^{i\omega t} \langle \left[j_i^x(t),\ j_j^x(0)\right] \rangle$
in the presence of fluctuations of the order parameter, amplitude $A_i$ and phase $\Phi_i$, and density $\delta n_i$ where $j^x$ stands for the current along the $x$ direction and $i,j$ are site indices. 
These corrections to the BdG results, evaluated using the random phase approximation~\cite{Supplementary,cea2014}, that includes vertex corrections required to restore gauge invariance, leads to 
\begin{align}
	\chi_{ij}\!\left(j^x,j^x\right) \!=\! \chi^0_{ij}\!\left(j^x,j^x\right) + \Lambda_{ip}\mathbb{V}_{pl} \left(\mathbb{I}_{3N\times 3N}\!-\!\chi^B \mathbb{V}\right)^{-1}_{ls}\bar{\Lambda}_{sj}.
	\label{eq.fullchi}
\end{align}
Here $\chi^0$ is the bare current-current correlation function, $\Lambda$ is the correlation function between current and one of the fluctuation components, $\chi^B$ is the bare mean-field susceptibility and $\mathbb{V}$ is the effective local interaction, defined by a $3 \times 3$ matrix in the fluctuation basis. Importantly,  all these quantities~\cite{Supplementary} can be expressed in terms of the parameters of the model and the output of the previous BdG calculation. 
Before we embark in the calculation of the conductivity, we aim to characterize collective excitations by investigating the spatial structure of these susceptibilities, $C^{ab}(r,\omega)=\langle\tilde{\chi}^{ab}(r,\omega)\rangle/\langle\tilde{\chi}^{ab}(0,\omega)\rangle$ where $r=|r_i-r_j|$, $\langle \ldots \rangle$ stands for spatial and disorder average, $a,b$ label the fluctuation channel, $\tilde\chi^{ab}$ is the block matrix of $\tilde{\chi}^B$(see supplemental information \cite{Supplementary} for details), and
\begin{align}
	\tilde{\chi}^B = \left(\mathbb{I}_{3N\times 3N}-\chi^B \mathbb{V}\right)^{-1}\chi^B. 
	\label{eq.RPA}
\end{align}
In the weak coupling limit, we shall see that in most cases, sub-gap weight in the conductivity is dominated by phase fluctuations $a=b=\Phi$, so we restrict to this channel,  
\begin{align}\label{cr}
	C(r,\omega) \equiv C^{\Phi\Phi}(r,\omega). 
\end{align}
Physically, it describes phase correlations in points of the sample separated by a distance $r$ after a perturbation of energy $\omega$. If $C(r,\omega)>0$ for $r \to \infty$, phase coherence holds. For our purposes, we define a {\it dephasing} length $\ell$, as the typical distance between a local maximum and a local minimum in $C(r,\omega)$. A necessary condition for the existence of phase collective excitations, the Goldstone mode, at a given energy $\omega$, is that $\ell < L$ otherwise phases are not sufficiently uncorrelated to become optically active.
\begin{figure}[t]
		\subfigure[]{\label{fig.pha_corU1}
			\includegraphics[width=4.cm]{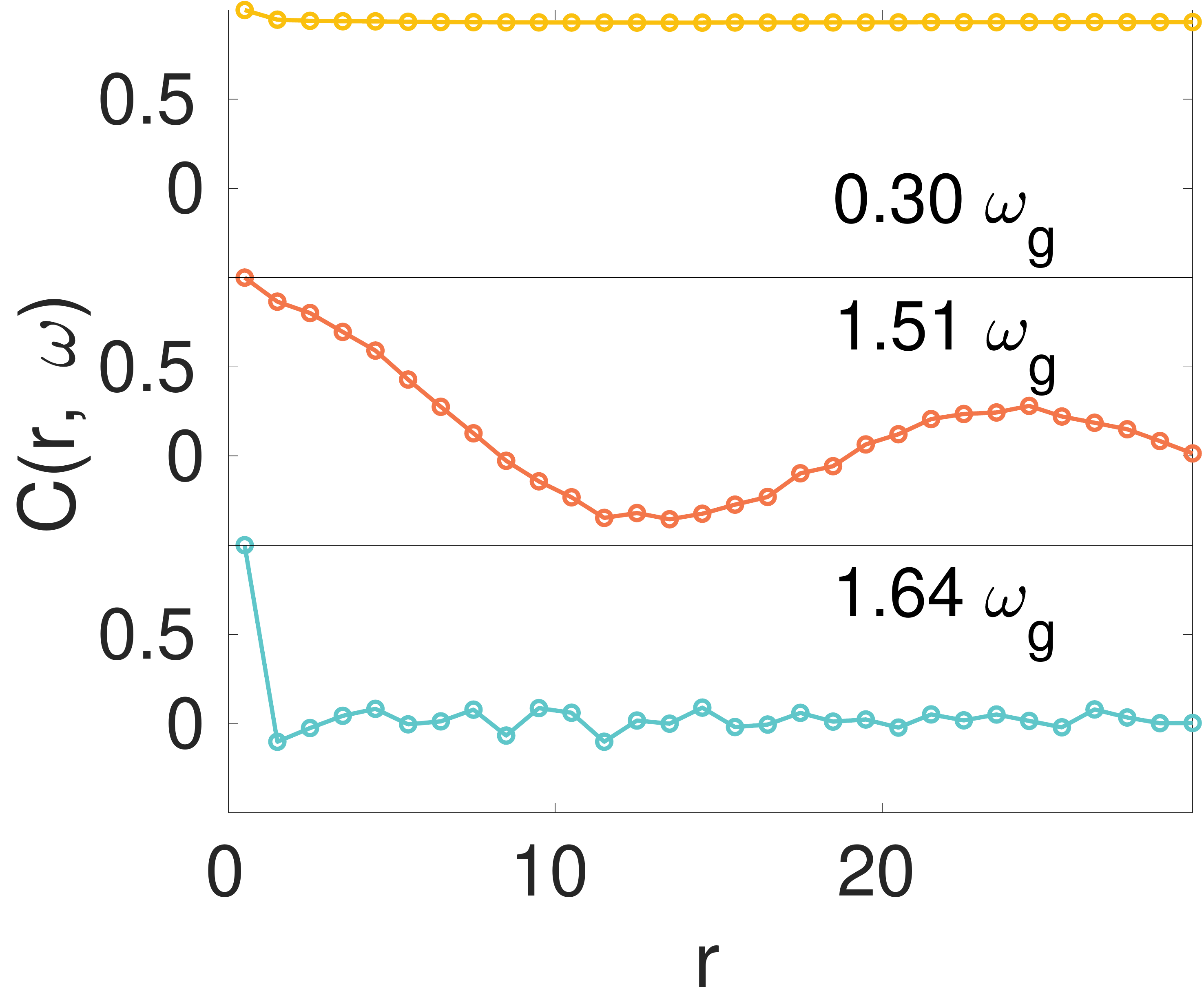}}
		\subfigure[]{\label{fig.pha_corU1_V0p5}
			\includegraphics[width=4.cm]{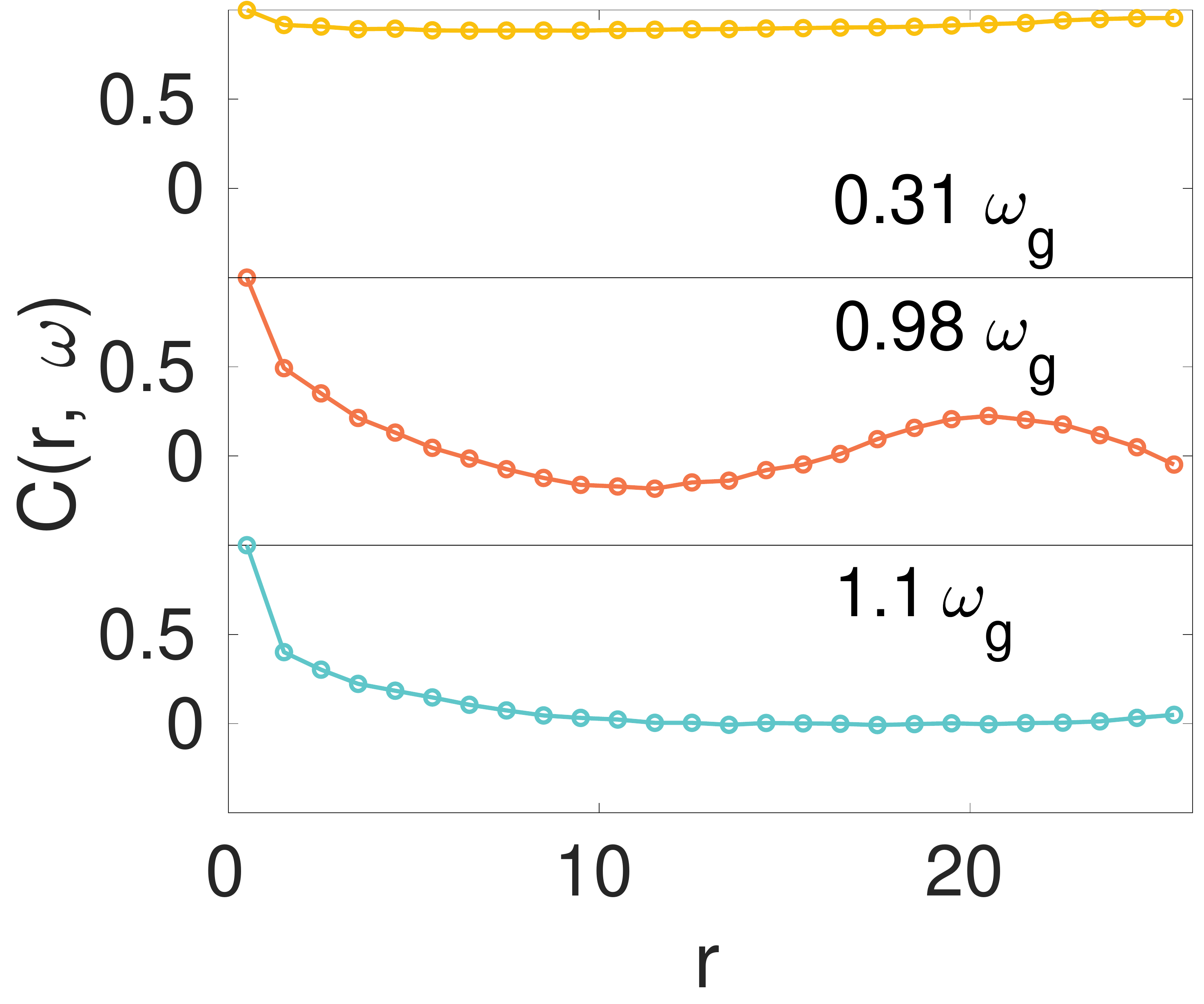}}
		\subfigure[]{\label{fig.pha_corU1_V1p5}
			\includegraphics[width=4.cm]{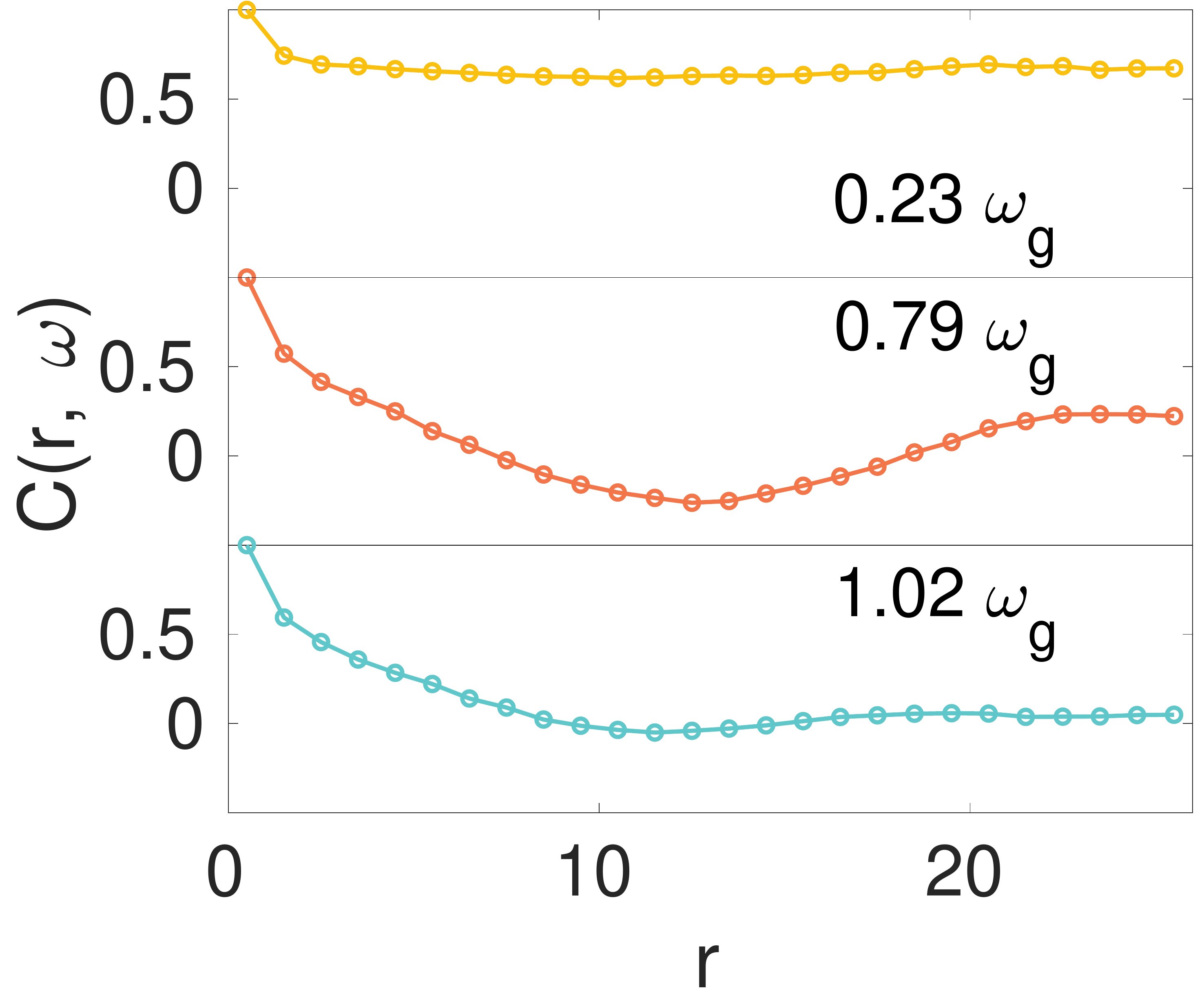}}
		\subfigure[]{\label{fig.pha_corU1_V3p0}
			\includegraphics[width=4.cm]{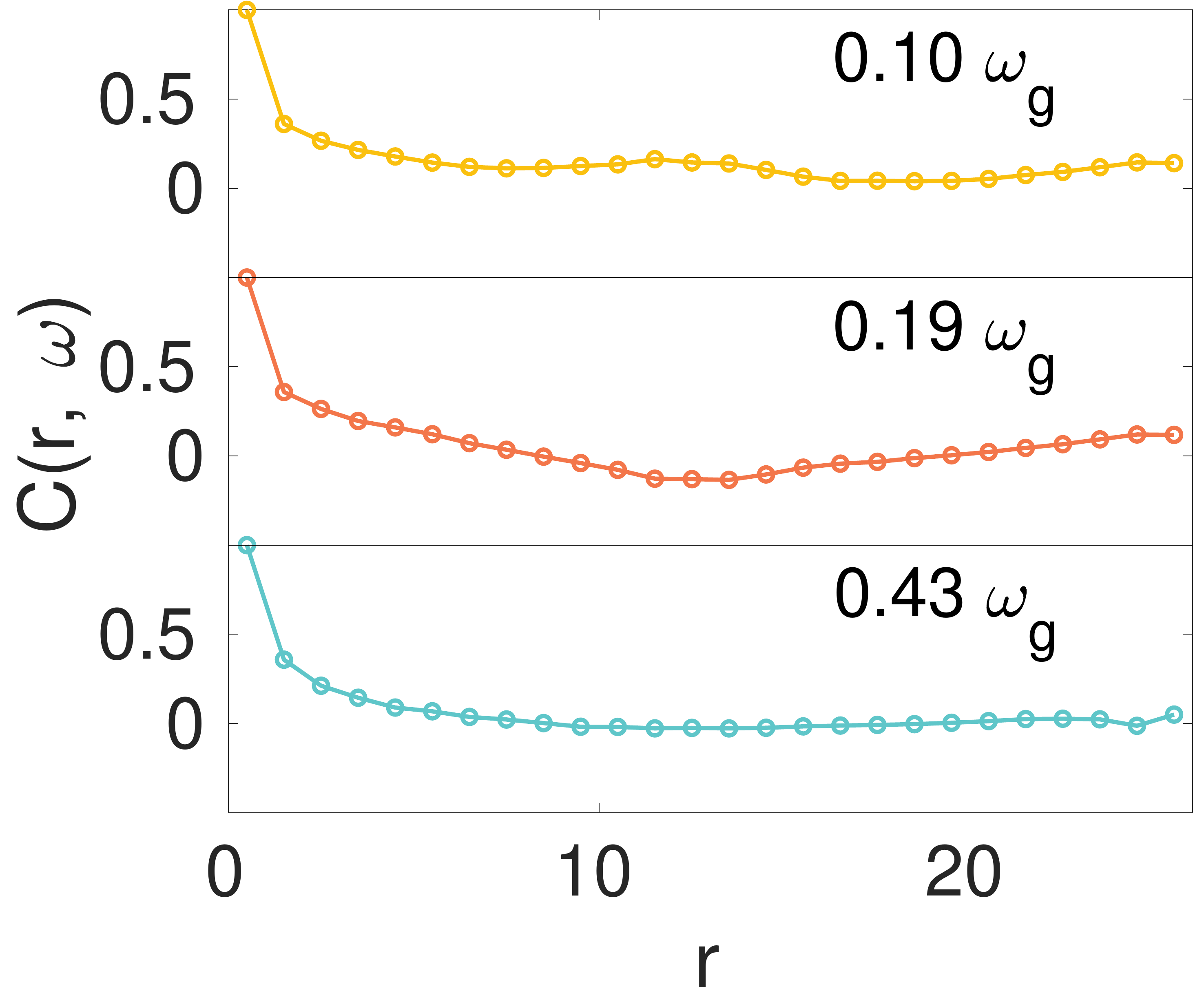}}
		\caption{$C(r,\omega)$ Eq.~(\ref{cr}) for different $\omega$ in units of the two-particle spectral gap $\omega_g$, $U=1$, $L=26$ (except for $V = 0$), $\langle n \rangle = 0.875$ and: \subref{fig.pha_corU1} $V=0$ ($L=30$), \subref{fig.pha_corU1_V0p5} $V=0.5$, \subref{fig.pha_corU1_V1p5} $V=1.5$, and \subref{fig.pha_corU1_V3p0} $V=3.0$. The observed oscillations, with $C(r,\omega)$ alternating sign, is a defining feature of collective modes. 
		} \label{Fig.pha_corU1}
\end{figure}
For clean or weak disorder, and small $\omega \ll$ two-particle spectral gap ($\omega_g$),
$C(r,\omega) \ge 0$ decays monotonously so phase fluctuations are still too correlated for a Goldstone mode to be observed. For very strong disorder, $C(r,\omega) \to 0$ quickly so no collective excitations can occur. For intermediate disorder, we expect that phases become sufficiently uncorrelated but still phase coherence can hold. This behavior is naturally related to oscillations in $C(r,\omega)$ that can become negative signaling phase fluctuations are strong enough that phases in distant points become anti-correlated. Qualitatively, the number of optically active regions is giving by the number of times that $C(r,\omega)$ switches sign (see \cite{Supplementary} for more details). If these features occur for $\omega <\omega_g$, the Goldstone mode is observed as a sub-gap excitation of the ac conductivity. 

Numerical results largely confirm this picture. In the clean or weak disorder region $V \le 0.5$ (Figs.~\ref{fig.pha_corU1}, \subref{fig.pha_corU1_V0p5}), oscillations around $0$ only occur in a narrow window of energies above $\omega_g$ ($\omega \sim \omega_g$ when $V=0.5$) and therefore are not relevant for the observation of the Goldstone mode in the conductivity that requires a well formed sub-gap peak. Phase coherence ($C(r,\omega) > 0$ for $r = L$) holds unless $\omega$ is not too large.

For sufficiently strong disorder $V = 1.5$ (Fig.~\ref{fig.pha_corU1_V1p5}), and small $\omega$, $C(r,\omega)$ decays rapidly to a constant positive value that indicates no optical activity. As $\omega$ approaches $\omega_g$ from below, we observe a much slower decay to negative value, that defines the dephasing length $\ell$, and indicates the presence of the sub-gap Goldstone mode. For even stronger disorder $V=3$ (Fig.~\ref{fig.pha_corU1_V3p0}), already in the insulating region, we observe similar features around $\omega \sim 0.2\omega_g$, inducing negative value in $C(r,\omega$).   

In order to find out whether these modes are measurable, we now turn to the calculation of the ac conductivity.  
The real part of the optical conductivity is closely related ~\cite{cea2014,cea2015,Seibold2015} to the susceptibility computed previously,
\begin{equation}
	\sigma(\omega) = \pi D_s\delta(\omega) + e^2{\rm Im} \frac{\chi(\omega)}{\omega} \label{sigma}
\end{equation} 
where $\rm Im$ stands for the imaginary part of $\chi(\omega) = 1/N\sum_{ij}\chi_{ij}(j^x,j^x)$, $e$ is the elementary charge, $D_s = e^2 [\langle -k_x \rangle + {\rm Re} \chi(\omega=0)]$ is the superfluid stiffness, and $\langle -k_x \rangle = 4\langle \sum_{n,i} v_n(i)v_n(i+\hat{x}) \rangle /N$ is the kinetic energy along the $x$ direction.
\begin{figure}
		\subfigure[]{\label{fig.Cond_U1_2}
			\includegraphics[width=4cm]{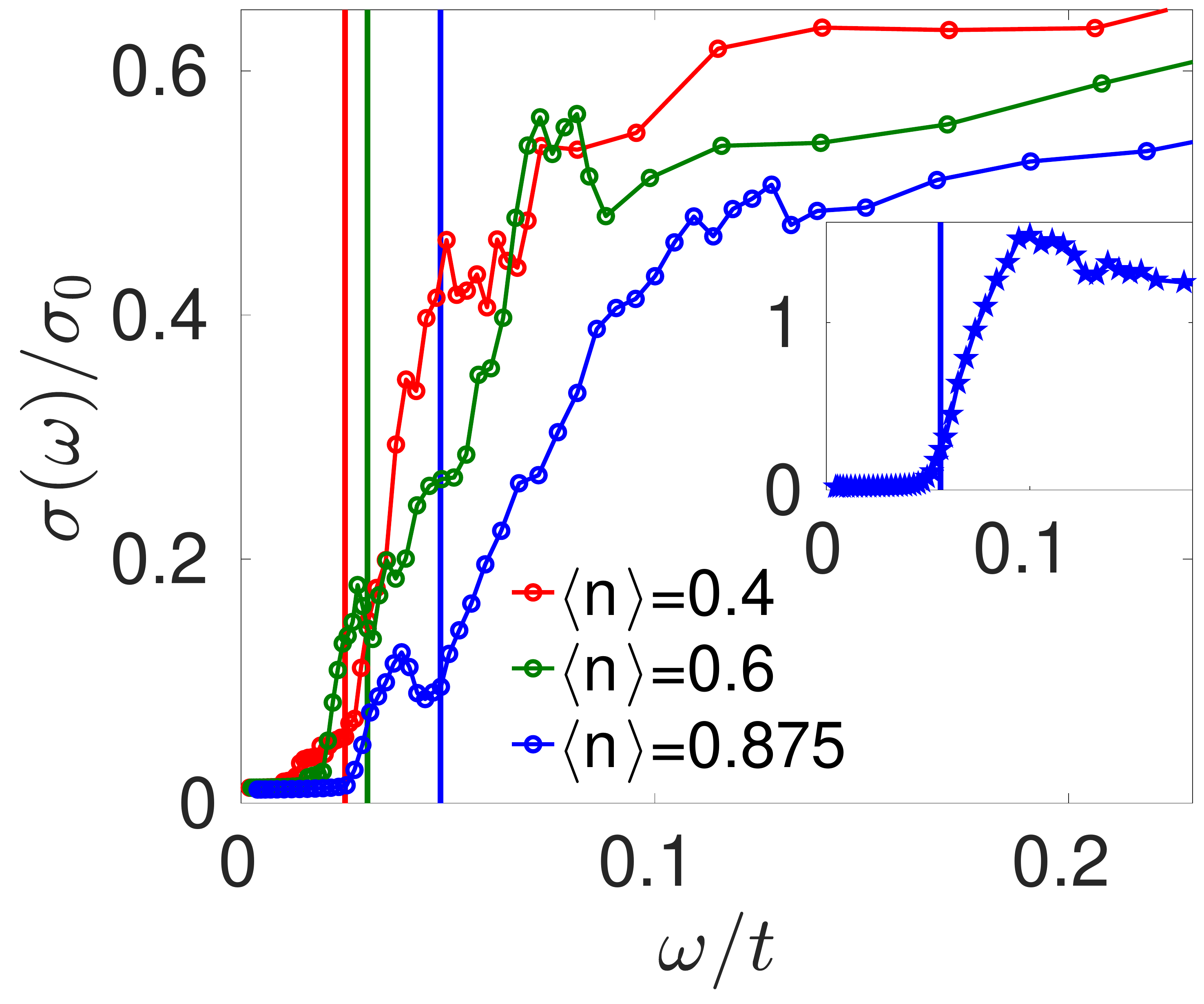}}
		\subfigure[]{\label{fig.Cond_U1_3}
			\includegraphics[width=4cm]{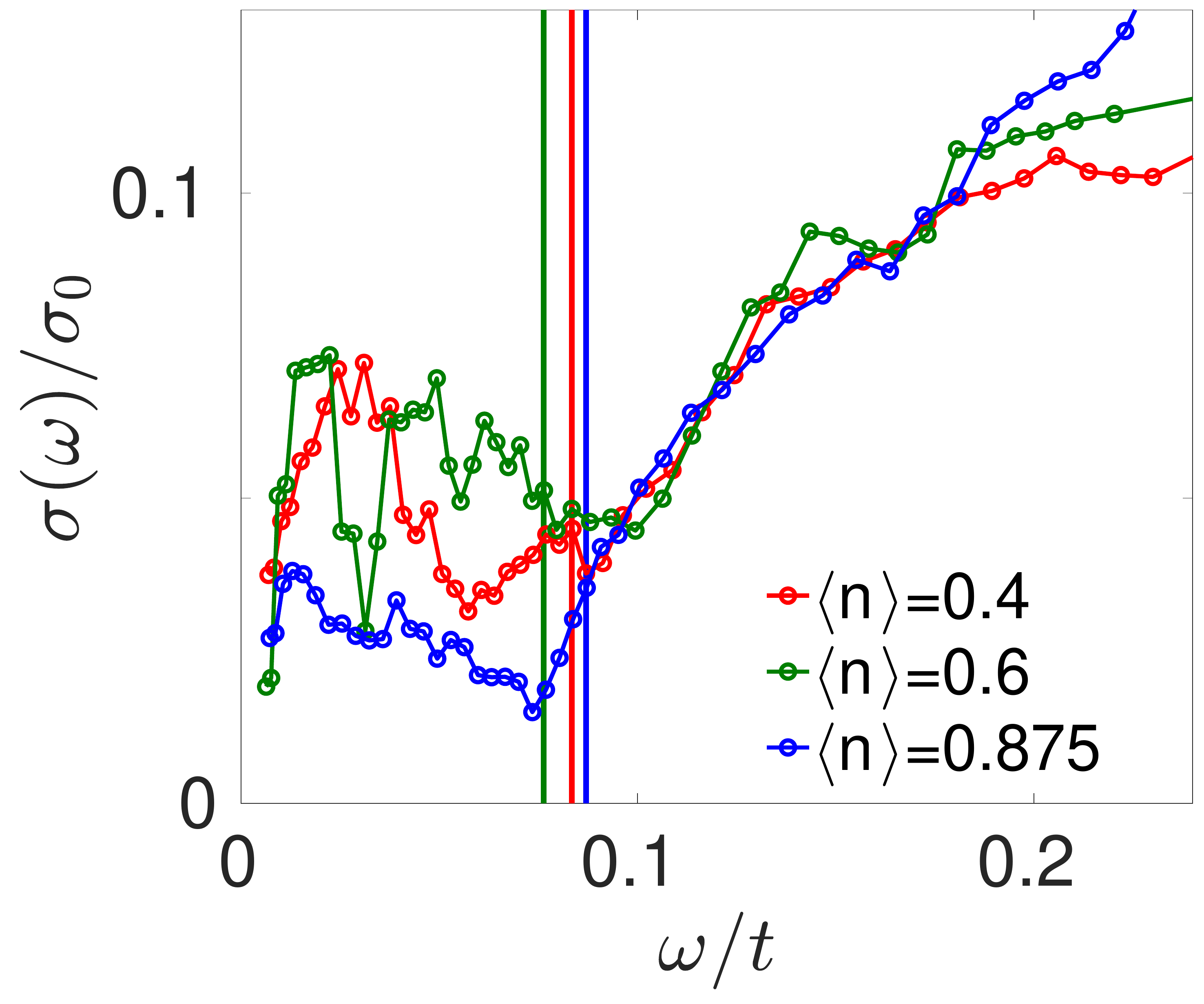}}
		\caption{Optical conductivity $\sigma(\omega)$ in units of $\sigma_0=\frac{e^2}{\hbar}$ for $U=1, L=26$ and different $V$ and $\langle n \rangle$. Left:  \subref{fig.Cond_U1_2} $V=1.5$ where the sub-gap excitation (Goldstone mode) starts to be observed. Inset: $U=1, \langle n \rangle =0.875, V=0.5$. No sub-gap collective excitation. Right:
			\subref{fig.Cond_U1_3} $V=3$, the sub-gap spectral weight broadens reaching very low frequencies. 
		} \label{Fig.Cond_U1}
\end{figure}
In Fig.~\ref{Fig.Cond_U1}, we depict the conductivity in the weak coupling region $U=1$ for different disorder strengths $V$. For $\langle n \rangle =0.4$ and $0.6$, size effects could be important when $V \leq 1$, so for $V=0.5$ we restrict ourselves to $\langle n \rangle = 0.875$, see inset of Fig.~\ref{fig.Cond_U1_2}, where this problem does not arise. Results are consistent with the previous calculation of $C(r,\omega)$. 
For weak disorder, we do not observe any clear sub-gap structure despite the fact that for $V \sim 1$ the order parameter is already strongly inhomogeneous. This is in contrast with the strong coupling limit \cite{cea2014,Supplementary} where sub-gap spectral weight is observed even for $V < 1$. Superficially, this seems surprising because strong coupling means a much larger order parameter. However, note that the existence of collective excitations depends on how correlated in space is the order parameter.  In weakly coupled superconductors, due to a larger coherence length, neighboring sites are more likely to be correlated which makes more difficult to become optically active.

We do observe a clear sub-gap weight related to collective excitations only for $V \gtrsim 1.5$. For $V \sim 1.5$, the sub-gap mode is peaked close to $\omega_g$ with no spectral weight elsewhere also in  agreement with $C(r,\omega)$ (see Fig.~\ref{fig.pha_corU1_V1p5}).
This indicates that only one or very few large domains become optically active. As $V$ increases, the typical length $\ell$ that controls the decay of $C(r,\omega)$ becomes smaller and more domains become optically active resulting in a broader spectrum, see Fig.~\ref{fig.Cond_U1_3} for $V=3$. The region $V>3$ (not shown) is similar to the disordered strong-coupling limit where no phase coherence holds and only isolated islands act like nano-antennas for the partial absorption of the electromagnetic radiation. 
\begin{figure}
		\subfigure[]{\label{fig.U1_cor2}
			\includegraphics[width=4.cm]{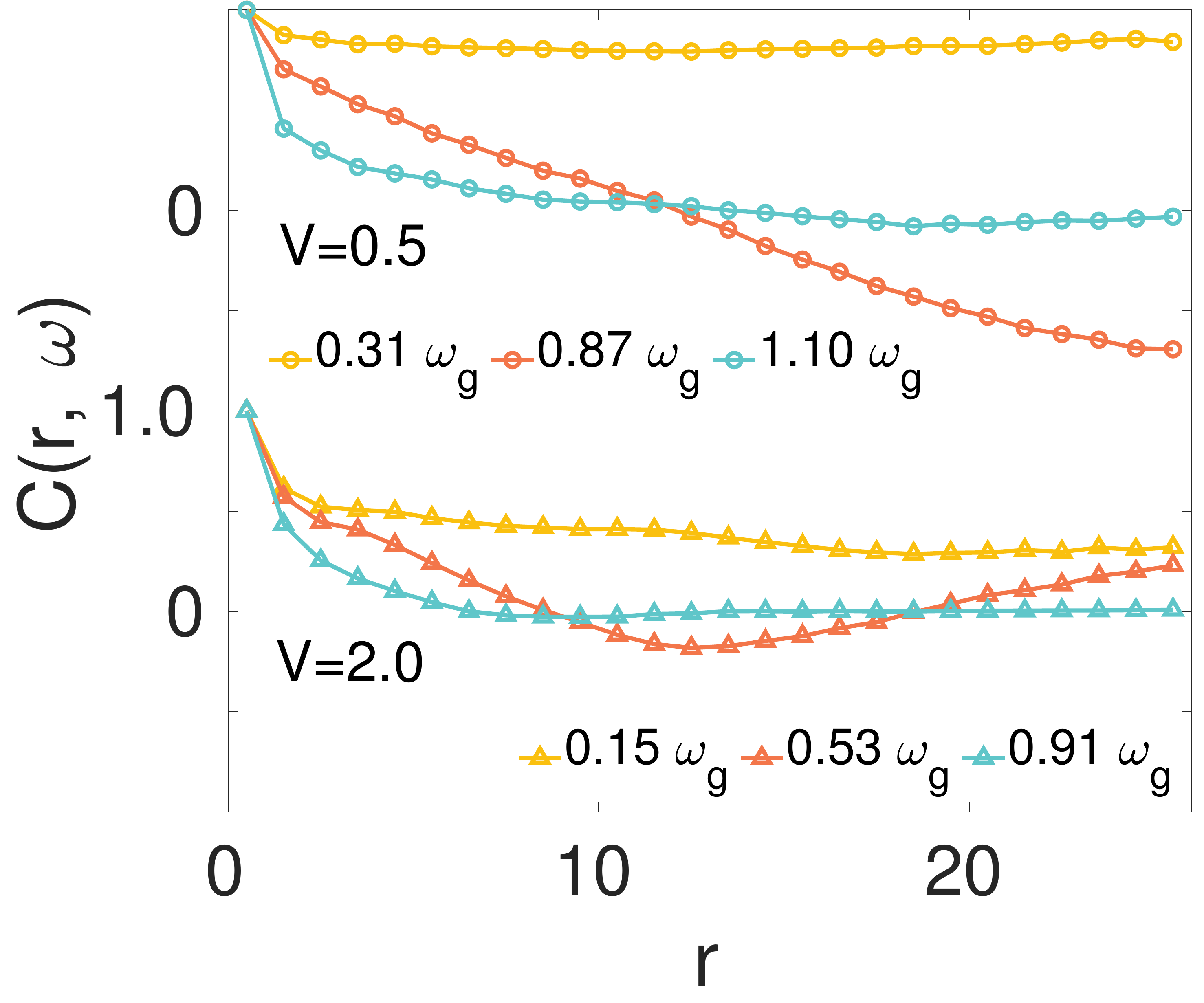}}
		\subfigure[]{\label{fig.U1_di2_L}
			\includegraphics[width=4.cm]{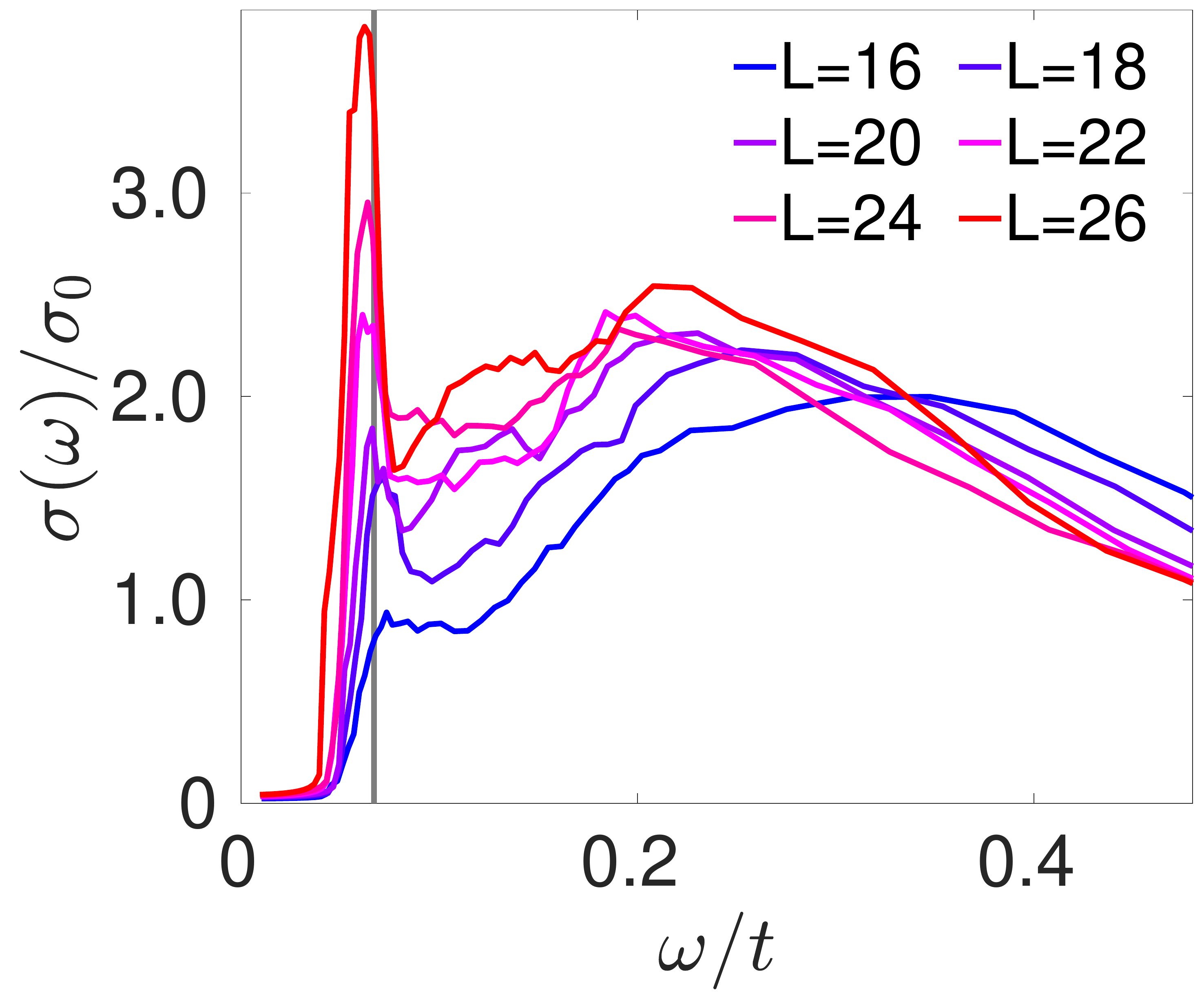}}
		\subfigure[]{\label{fig.U1_di3_L}
			\includegraphics[width=4.cm]{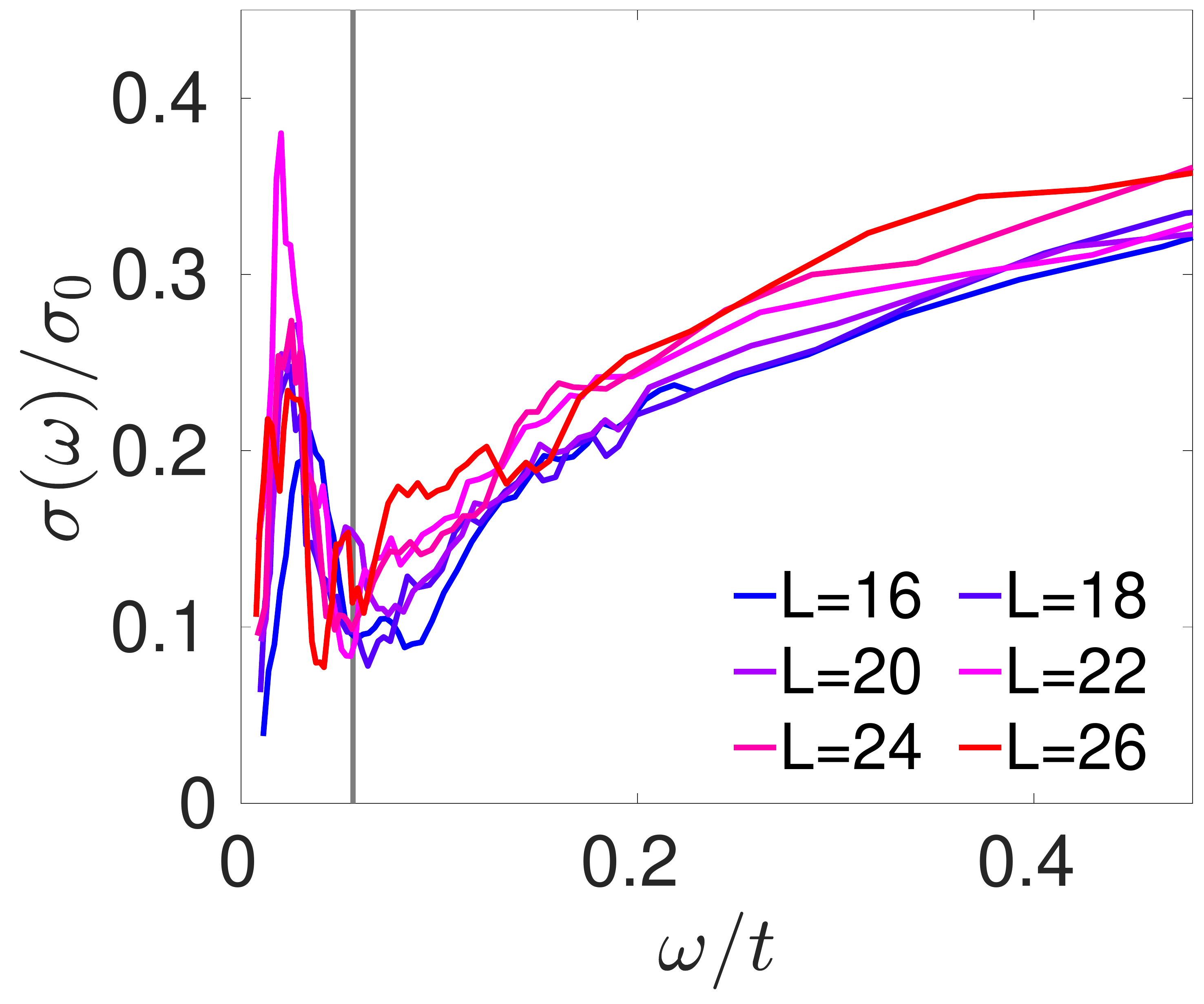}}
		\subfigure[]{\label{fig.U1_di1_L}
			\includegraphics[width=4.cm]{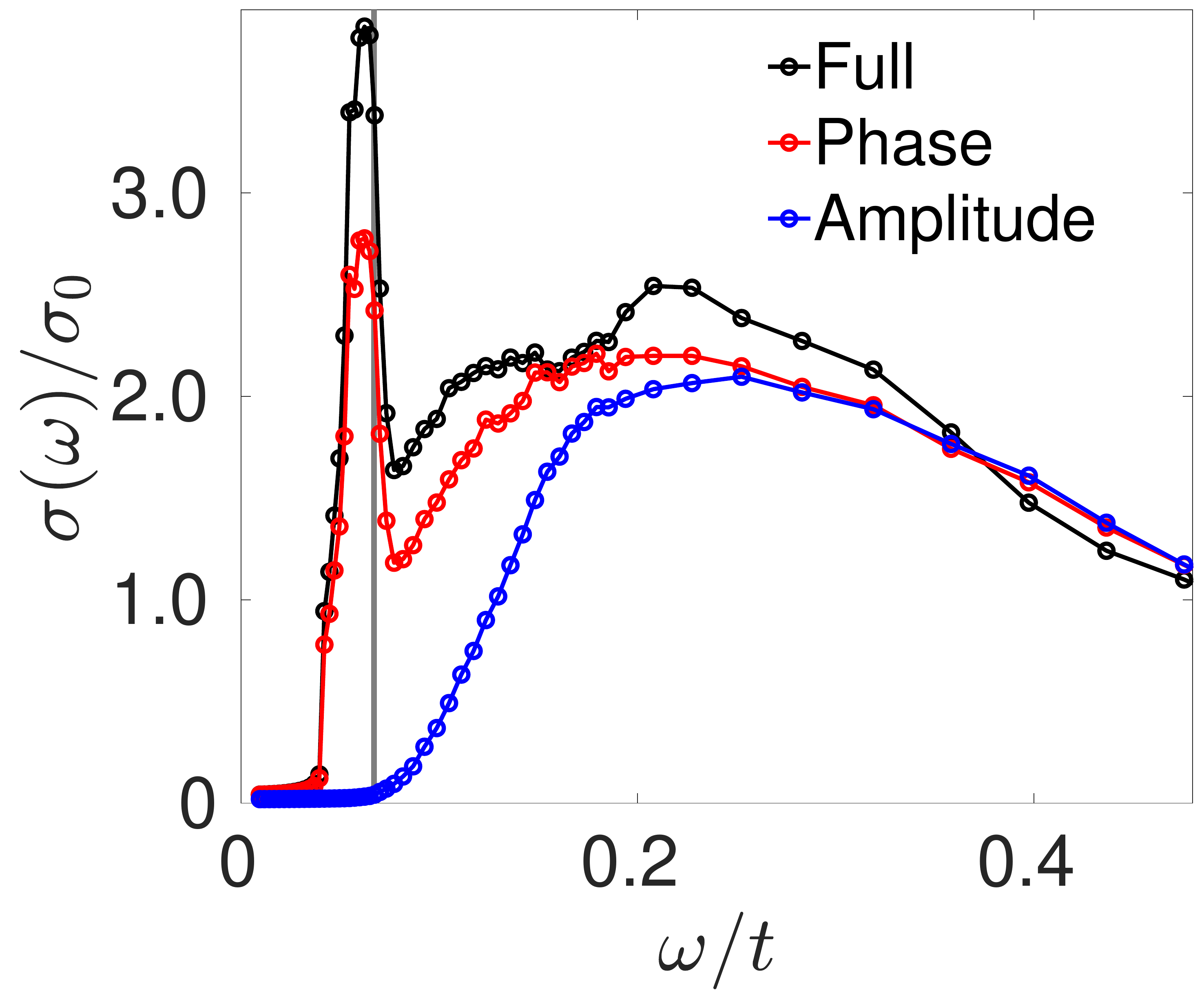}}
		\caption{Conductivity and $C(r,\omega)$ Eq.~(\ref{cr}) for Dirichlet boundary conditions. \subref{fig.U1_cor2}: $C(r,\omega)$ for $U=1, L=26, \langle n \rangle =0.875$. Upper: $V=0.5$. Lower: $V=2$. \subref{fig.U1_di2_L}: $\sigma(\omega)$ for different sizes $L$, with parameters of \subref{fig.U1_di1_L}. We observe a sub-gap collective excitation only for $L \geq 18$. This is the typical length $\ell$ for some substantial dephasing to occur so that the region becomes optically active. \subref{fig.U1_di3_L}: $\sigma(\omega)$ for different sizes $L$, with parameters of \subref{fig.U1_cor2}. We observe sub-gap weight at similar energies for all sizes. This is fully consistent with $C(r,\omega)$ in \subref{fig.U1_cor2}. \subref{fig.U1_di1_L}: $\sigma(\omega)$ for $U=1, V = 0.5, L=26$, and $ \langle n \rangle =0.875$. Phase fluctuations control the sub-gap weight which is interpreted as the Goldstone mode.}\label{Fig.U1_cond_L}
\end{figure}
As a further confirmation of the relation between collective excitations and the existence of a {\it dephasing} length $\ell$, not related to isolated islands, we compute the conductivity for different sizes $L$  using Dirichlet boundary conditions that enhance finite size effects as it is imposed that the order parameter vanishes at the boundary. The idea is that for a given disorder strength, we will observe collective excitations around $\omega_g$ only if $\ell < L$. For smaller sizes, phases are not sufficiently uncorrelated for collective excitations to occur below $\omega_g$.   
Results depicted in Figs.~\ref{fig.U1_di2_L},~\subref{fig.U1_di3_L} confirm that for not too large $V$, a sub-gap peak requires a minimum system size. Moreover, $\sigma(\omega)$ and $C(r,\omega)$, see Fig.~\ref{fig.U1_cor2}, are not qualitatively altered by the change in boundary conditions but, as was expected, finite size effects are enhanced so the minimum disorder $V \sim 0.5$ at which collective excitation occur is weaker than for periodic boundary conditions $V\sim1.5$. This could help the experimental observation of collective excitations in sub-micron flakes \cite{verdu2018} of disordered superconductors.

We have referred to the sub-gap spectral weight as the Goldstone mode in several occasions but, so far, we have not provided explicit evidence that this is the case. This is remedied in Fig.~\ref{fig.U1_di1_L}, where it is shown the conductivity, including full vertex corrections and still using Diricihlet boundary conditions, is qualitatively similar if only phase fluctuations are considered.

For the experimental confirmation of these results, it is important that the explored parameters $U=1$, $\langle n \rangle=0.4, 0.6, 0.875$ describe weakly-coupled materials like Al, Sn, Pb or Nb. A simple calculation of the coherence length $\xi_0$ based on $\Delta_0$, and standard BCS relations, yields that our results apply to materials with $\xi_0 \lesssim 500$nm which, though short for Al, cover most weakly-coupled materials. We stress that in the relevant $V \geq 1$ region, finite size effects for all $\langle n \rangle$ are negligible. Another important issue is whether the Coulomb interactions, neglected here, alter qualitatively our main findings. In Ref.~\cite{cea2014} it was argued that, at least for the conductivity, this is not the case. We also believe that, at least for not very strong disorder, long range Coulomb interaction is heavily suppressed and therefore it should not alter substantially the Goldstone mode typical frequency. It is an open question to what extent other features are quantitatively influenced by residual Coulomb interactions.

In summary, we have shown that sub-gap excitations in the optical conductivity can be observed in weakly-coupled disordered superconductor provided that spatial inhomogeneities of the order parameter are sufficiently strong so that the typical length of decay of phase fluctuations is smaller than the system size. 
Therefore, unlike strongly coupled superconductors \cite{cea2014}, collective excitations can coexist with a finite supercurrent and do not require the existence of isolated superconducting islands acting like nano-antennas. 
We expect our results stimulate experimental interest in this problem that could lead to a full characterization of collective modes in disordered metallic superconductors.  

\acknowledgments{ 
	B.F. and A.M.G.G. acknowledge financial support from a Shanghai talent program, from the National Natural Science Foundation China (NSFC) (Grant No. 11874259) and from the National Key R\&D Program of China (Project ID: 2019YFA0308603). A.M.G.G. thanks valuable conversations with Lara Benfatto. A.S. and B.F. thank illuminating conversations with Goetz Seibold that, among other things, help solve a technical problem with the code.}


\bibliographystyle{unsrt}
\bibliography{library1}
\onecolumngrid
\clearpage

\setcounter{affil}{0}
\renewcommand{\thefigure}{S\arabic{figure}}
\setcounter{figure}{0}
\renewcommand{\theequation}{S\arabic{equation}}
\setcounter{equation}{0}
\renewcommand\thesection{S\arabic{section}}
\setcounter{section}{0}

\title{Supplemental material for ``Characterization of collective excitations in weakly-coupled disordered superconductors"}
\label{SI}


\begin{abstract}
\noindent We present the technical details involving the calculation of the full current-current correlator. Next we study the behavior of the collective modes in the clean limit, as well as in presence of disorder. Finally we discuss the amplitude and phase fluctuation correlation functions and the optical conductivity.
\end{abstract}
\maketitle
\onecolumngrid
\section{I.\,\,\,\,\,\,     Current-current correlator in disordered superconductors}
We consider a disordered attractive Hubbard model on a square lattice in presence of onsite random potential $V_i$ ($V_i\in [-V,V]$). The model Hamiltonian is given by,
\begin{equation}
H = -t\sum_{\langle ij\rangle\sigma} c^\dagger_{i\sigma} c_{j\sigma} - U\sum_i n_{i\uparrow}n_{i\downarrow} + \sum_i V_i n_i
\end{equation} 
where $t$ is the hopping amplitude between two nearest neighbors, and $U$ is the attractive interaction responsible for the Cooper pairing. We invoke an inhomogeneous mean-field theory (Bogoliubov de-Gennes theory) with two mean-field parameters: local superconducting order parameter $\Delta(r_i)$ and local density $n(r_i)$~\cite{Ghosal2001, ghosal1998}.  We then use the following Bogoliubov transformation,
\begin{equation}
c_{i\sigma} = \sum_n \lt(u_n(i)\gamma_{n\sigma} - \sigma v^*_n(i)\gamma^\dagger_{n\bar\sigma}\rt)
\end{equation}
which diagonalizes the effective mean-field Hamiltonianin in fermionic quasi-particle basis ($\gamma$),
\begin{equation}
H_{MF} = \sum_{n\sigma} E_n \gamma^\dagger_{n\sigma}\gamma_{n\sigma},
\end{equation}
where $n$ runs over the positive eigenvalues i.e. $E_n>0$.

Next we study the effect of disorder on the optical response of the system. The dynamical correlation function is defind as \cite{cea2014}
\begin{equation}
\chi_{ij}(\phi,\phi')=-i \int dt e^{i\omega t} \langle \left[\phi_i(t),\ \phi'_j(0)\right] \rangle
\label{eq.cor_fun}
\end{equation} 
where $\phi$ corresponds to the fluctuation components and the current operators, which are given by \cite{cea2014}
\begin{equation}
\begin{aligned}
&\delta\Delta_i &=\ &c_{i\downarrow}c_{i\uparrow} - \langle c_{i\downarrow}c_{i\uparrow} \rangle \\
&\delta\Delta_i^\dagger &=\ &c_{i\uparrow}^\dagger c_{i\downarrow}^\dagger - \langle c_{i\uparrow}^\dagger c_{i\downarrow}^\dagger \rangle \\
&\delta n_i &=\ &\sum_{\sigma} \left( c_{i\sigma}^\dagger c_{i\sigma} - \langle c_{i\sigma}^\dagger c_{i\sigma} \rangle \right) \\
&j_i^\alpha &=\ &it \sum_{\sigma} \left( c_{i+\alpha,\sigma}^\dagger c_{i\sigma} - c_{i\sigma}^\dagger c_{i+\alpha,\sigma} \right). \\
\end{aligned}
\label{eq.fluct_curr}
\end{equation} 
Here $\delta\Delta_i$ is the fluctuation in local superconducting order parameter, and $\delta n_i$ is the fluctuation in local density. $\langle\cdots\rangle$ corresponds to the expectation value of the operator in the inhomogeneous BdG eigenstate. The amplitude fluctuation $A_i$ and the phase fluctuation $\Phi_i$ of the superconducting order parameter are given by 
\begin{align}
A_i=(\delta\Delta_i+\delta\Delta_i^\dagger)/\sqrt{2} \label{eq.amp}\\
\Phi_i=i(\delta\Delta_i-\delta\Delta_i^\dagger)/\sqrt{2}
\label{eq.pha}
\end{align}
Note that all the eigenvectors $(u_n, v_n)$ in our case are real, and hence we can express the dynamical correlation functions in terms of $u_n, v_n$ and $E_n$ only. Now we present the detailed formulae for different correlation functions at zero temperature.

The bare current-current correlation function is given by~\cite{Ghosal2001},
\begin{eqnarray}
\nonumber \chi^0_{ij}(j^x,j^x) 
&= -2t^2\sum_{nm}\left(v_{m}(j+\hat x)u_{n}(j) + v_{n}(j+\hat x)u_{m}(j)\right) \left({\frac{u_n(i+\hat x)v_m(i) - u_n(i)v_m(i+\hat x)}{i\omega_p+E_n+E_m} } - {\frac{v_n(i)u_m(i+\hat x) - v_n(i+\hat x)u_{m}(i)}{i\omega_p-E_n-E_m} }\right) \no\\ 
&~~~~~~~~~~~~~~~~~~~~~~~~~~~~~~~~~~~~~~~~~~~~~~~~~~~~~~~~~~~
+u\leftrightarrow v
\label{eq.bare_jj}
\end{eqnarray}
where $i\omega_p$ is the Bosonic Matsubara frequency, given by $\omega_p=2\pi p/\beta$. \\

The vertex correction to the bare current-current correlation function can be calculated by introducing the following correlations. The correlation functions between the current operator and the pair fluctuations ($\delta\Delta^\dagger,\delta\Delta$), or the charge density fluctuations ($\delta n$) are given by,
\begin{eqnarray}
\chi_{ij}(j^x,\delta\Delta) &=& 2it \sum_{nm} 
{\left(u_n(i+\hat x)v_m(i) - u_n(i)v_m(i+\hat x)\right)} \left( {u_{m}(j)u_{n}(j) \over i\omega_p +E_n+E_m} - {v_{m}(j)v_{n}(j) \over i\omega_p -E_n-E_m} \right) \\
\chi_{ij}(j^x,\delta\Delta^\dagger) &=& -2it \sum_{nm} {\left(u_n(i+\hat x)v_m(i)-u_n(i)v_m(i+\hat x)\right)} \left( { v_{m}(j)v_{n}(j) \over i\omega_p +E_n+E_m} - {u_{m}(j)u_{n}(j) \over i\omega_p -E_n-E_m} \right) \\
\chi_{ij}(j^x,\delta n) &=&  2it \sum_{nm} 
-{\left(u_n(i+\hat x)v_m(i)-u_n(i)v_m(i+\hat x)\right) \left(v_{m}(j)u_{n}(j) + v_{n}(j)u_{m}(j)\right) \over i\omega_p +E_n+E_m} \nonumber\\
&& ~~~~~~~~~~~~~~~~~~~~~~~~ + {\left(v_n(i+\hat x)u_m(i)-v_n(i)u_m(i+\hat x)\right) \left(u_{m}(j)v_{n}(j)+u_{n}(j)v_{m}(j)\right)  \over i\omega_p-E_n-E_m}
\end{eqnarray}

The bare correlation functions between pair fluctuations ($\delta\Delta^\dagger,\delta\Delta$) and charge density fluctuations ($\delta n$) are given by,
\begin{align}
\chi_{ij}(\delta\Delta,\delta\Delta) &= \sum_{nm} -{u_n(i)u_m(i) v_m(j)v_n(j) \over i\omega_p-E_n-E_m}
+ {v_n(i)v_m(i) u_m(j)u_n(j) \over i\omega_p+E_n+E_m} \\
\chi_{ij}(\delta\Delta,\delta\Delta^\dagger) &= \sum_{nm} {u_n(i)u_m(i) u_m(j)u_n(j) \over i\omega_p-E_n-E_m}
- {v_n(i)v_m(i) v_m(j)v_n(j) \over i\omega_p+E_n+E_m} \\
\chi_{ij}(\delta\Delta^\dagger,\delta\Delta) &= \sum_{nm} {v_n(i)v_m(i) v_m(j)v_n(j) \over i\omega_p-E_n-E_m}
- {u_n(i)u_m(i) u_m(j)u_n(j) \over i\omega_p+E_n+E_m}\\
\chi_{ij}(\delta\Delta,\delta n) &= -2 \sum_{nm}  {u_n(i)u_m(i)u_{m}(j)v_{n}(j) \over i\omega_p-E_n-E_m}
+ {v_n(i)v_m(i) v_{m}(j)u_{n}(j) \over i\omega_p+E_n+E_m} \\
\chi_{ij}(\delta n,\delta \Delta) &= 2\sum_{nm} {u_n(i)v_m(i) u_{m}(j)u_{n}(j) \over i\omega_p+E_n+E_m}
+ {v_n(i)u_m(i) v_{m}(j)v_{n}(j) \over i\omega_p-E_n-E_m} \\
\chi_{ij}(\delta n,\delta n) &= 2\sum_{nm} 
{ \left(v_{m}(j)u_{n}(j) + v_{n}(j)u_{m}(j)\right)} \left( -{u_n(i)v_m(i)\over i\omega_p+E_n+E_m} +  {v_m(i)u_n(i)\over i\omega_p-E_n-E_m} \right)
\end{align}
\indent  The remaining correlation functions can be obtained using symmetry,
\begin{eqnarray}
\chi_{ij}(\Delta,j^x) &= &-\chi_{ji}(j^x,\Delta^\dagger) \no\\
\chi_{ij}(\Delta^\dagger,j^x) &= &-\chi_{ji}(j^x,\Delta) \no\\
\chi_{ij}(n,j^x) &= &-\chi_{ji}(j^x,n) \\
\chi_{ij}(\Delta^\dagger,\Delta^\dagger) &= &\chi_{ji}(\Delta,\Delta) \no\\
\chi_{ij}(\Delta^\dagger,n) &= &\chi_{ji}(n,\Delta) \no\\
\chi_{ij}(n,\Delta^\dagger) &= &\chi_{ji}(\Delta,n) \no
\label{eq.G1}
\end{eqnarray}

With the definitions of amplitude and phase fluctuations given in Eqn.~\eqref{eq.amp} and ~\eqref{eq.pha}, we can write the correlation functions between current and amplitude or phase fluctuations in the following way,
\begin{eqnarray}
\chi_{ij}(j^x,A) &=\frac{1}{\sqrt{2}} \left( \chi_{ij}(j^x,\delta\Delta) + \chi_{ij}(j^x,\delta\Delta^\dagger) \right)\\
\chi_{ij}(A,j^x) &=\frac{1}{\sqrt{2}} \left( \chi_{ij}(\delta\Delta,j^x) + \chi_{ij}(\delta\Delta^\dagger,j^x) \right)\\
\chi_{ij}(j^x,\Phi) &=\frac{i}{\sqrt{2}} \left( \chi_{ij}(j^x,\delta\Delta) - \chi_{ij}(j^x,\delta\Delta^\dagger) \right)\\
\chi_{ij}(\Phi,j^x) &=\frac{i}{\sqrt{2}} \left( \chi_{ij}(\delta\Delta,j^x) - \chi_{ij}(\delta\Delta^\dagger,j^x) \right),
\label{eq.4}
\end{eqnarray} 
and the correlation functions between amplitude fluctuation, phase fluctuation and density fluctuation are 
\begin{eqnarray}
&\chi_{ij}(A,A)&= \frac{1}{2} \left[ \chi_{ij}(\delta \Delta,\delta \Delta) + \chi_{ij}(\delta \Delta,\delta \Delta^\dagger) + \chi_{ij}(\delta \Delta^\dagger,\delta \Delta) + \chi_{ij}(\delta \Delta^\dagger,\delta \Delta^\dagger) \right]\\
&\chi_{ij}(A,\Phi)&= \frac{i}{2} \left[ \chi_{ij}(\delta \Delta,\delta \Delta) - \chi_{ij}(\delta \Delta,\delta \Delta^\dagger) + \chi_{ij}(\delta \Delta^\dagger,\delta \Delta) - \chi_{ij}(\delta \Delta^\dagger,\delta \Delta^\dagger) \right]\\
&\chi_{ij}(A,\delta n)&= \frac{1}{\sqrt{2}} \left[ \chi_{ij}(\delta \Delta,\delta n) + \chi_{ij}(\delta \Delta^\dagger,\delta n) \right]\\
&\chi_{ij}(\Phi,A)&= \frac{i}{2} \left[ \chi_{ij}(\delta \Delta,\delta \Delta) + \chi_{ij}(\delta \Delta,\delta \Delta^\dagger) - \chi_{ij}(\delta \Delta^\dagger,\delta \Delta) - \chi_{ij}(\delta \Delta^\dagger,\delta \Delta^\dagger) \right]\\
&\chi_{ij}(\Phi,\Phi)&=\frac{i^2}{2} \left[ \chi_{ij}(\delta \Delta,\delta \Delta) - \chi_{ij}(\delta \Delta,\delta \Delta^\dagger) - \chi_{ij}(\delta \Delta^\dagger,\delta \Delta) + \chi_{ij}(\delta \Delta^\dagger,\delta \Delta^\dagger) \right]\\
&\chi_{ij}(\Phi,\delta n)&= \frac{i}{\sqrt{2}} \left[ \chi_{ij}(\delta \Delta,\delta n) - \chi_{ij}(\delta \Delta^\dagger,\delta n) \right]\\
&\chi_{ij}(\delta n,A)&= \frac{1}{\sqrt{2}} \left[ \chi_{ij}(\delta n,\delta \Delta) + \chi_{ij}(\delta n,\delta \Delta^\dagger) \right]\\
&\chi_{ij}(\delta n,\Phi)&= \frac{i}{\sqrt{2}} \left[ \chi_{ij}(\delta n,\delta \Delta) - \chi_{ij}(\delta n,\delta \Delta^\dagger) \right]
\label{eq.5}
\end{eqnarray} 
\indent The full gauge invariant current-current correlation function (including the vertex corrections) is given by,
\begin{align}
\chi_{ij}\left(j^x,j^x\right) = \chi^0_{ij}\left(j^x,j^x\right) + \Lambda_{ip}\mathbb{V}_{pl} \left(\mathbb{I}_{3N\times 3N}-\chi^B \mathbb{V}\right)^{-1}_{ls}\bar{\Lambda}_{sj}
\label{eq.fullchi_SM}
\end{align}
where $\chi^0$ is the bare current-current correlation function, and $\Lambda$ couples the current with one of the fluctuation components. We note that we have three possible types of fluctuations i.e. $A$, $\Phi$ and $\delta n$ which correspond to amplitude, phase and charge density fluctuations respectively,
\begin{align}
\Lambda&=\left(~\chi(j^x,A)~~\chi(j^x,\Phi)~~\chi(j^x,\delta n)~\right)\\
\bar{\Lambda}&=\left(~\chi(A,j^x)~~\chi(\Phi,j^x)~~\chi(\delta n,j^x)~\right)^T.
\end{align}
$\chi^B$ is the bare mean-field susceptibility and $\mathbb V$ is the effective local interaction, defined by $3\times3$ matrices in the basis of fluctuations:
\begin{equation}
\chi^B = 
\left(\begin{array}{ccc}
\chi^{AA} & \chi^{A\Phi} & \chi^{A\delta n} \\
\chi^{\Phi A} & \chi^{\Phi \Phi} & \chi^{\Phi \delta n} \\
\chi^{\delta n A} & \chi^{\delta n\Phi} & \chi^{\delta n\delta n}
\end{array}\right)
\end{equation}
and 
\begin{equation}
\mathbb{V} = 
\left(\begin{array}{ccc}
-|U| & 0 & 0 \\
0 & -|U| & 0 \\
0 & 0 & -|U|/2
\end{array}\right)
\end{equation}
Therefore, in Eqn.~\eqref{eq.fullchi_SM}, for example, $\chi^B(A,A)=\chi^{AA}$, $\mathbb{V}^A=-|U|\mathbb{I_{N\times N}}$ and so on ... Note that all of them are $N\times N$ matrices in real space.\\
\begin{figure}[H]
	\begin{center}
		\subfigure[Amplitude]{\label{fig.amp_U5}
			\includegraphics[width=6.cm]{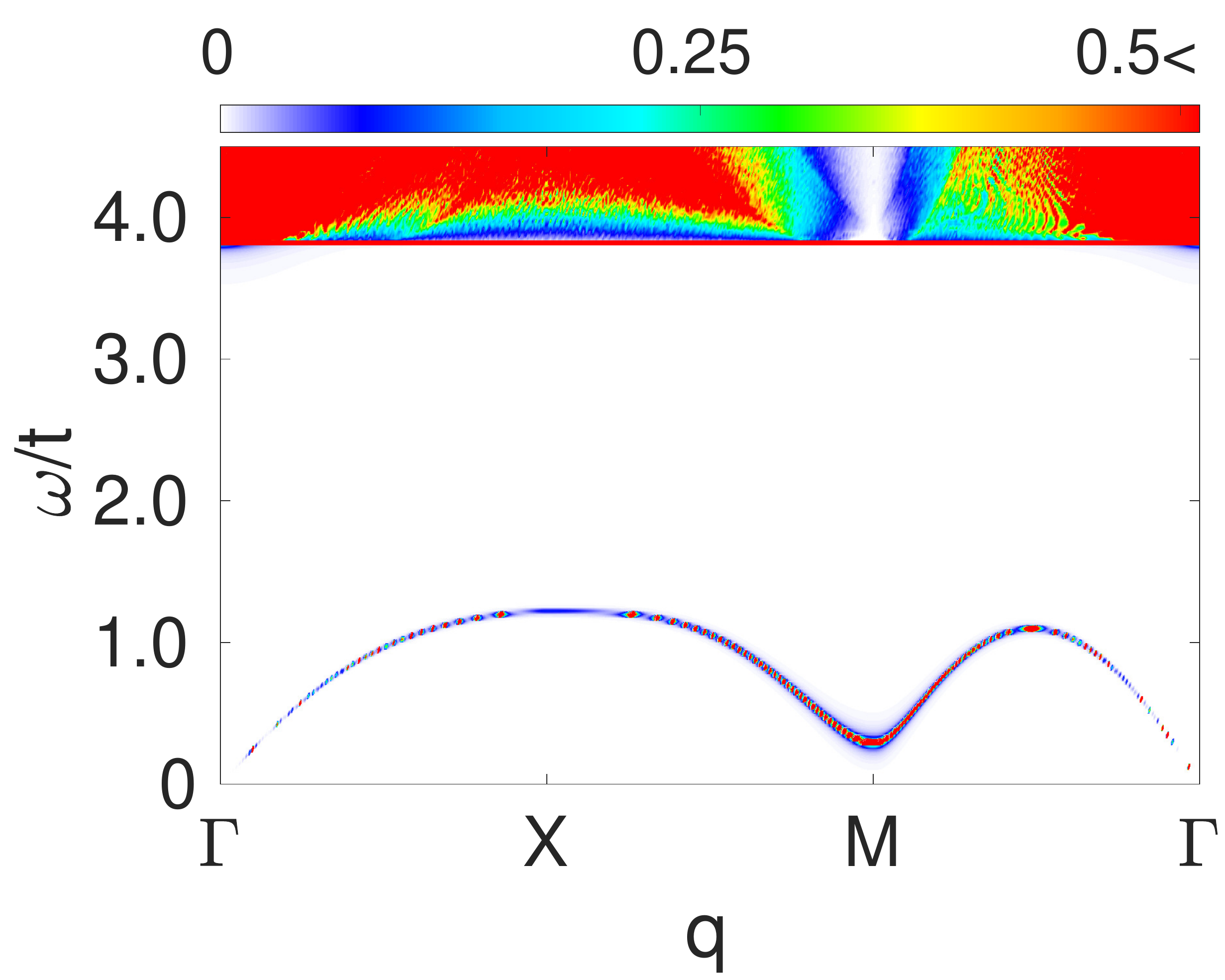}}
		\subfigure[Phase]{\label{fig.pha_U5}
			\includegraphics[width=6.cm]{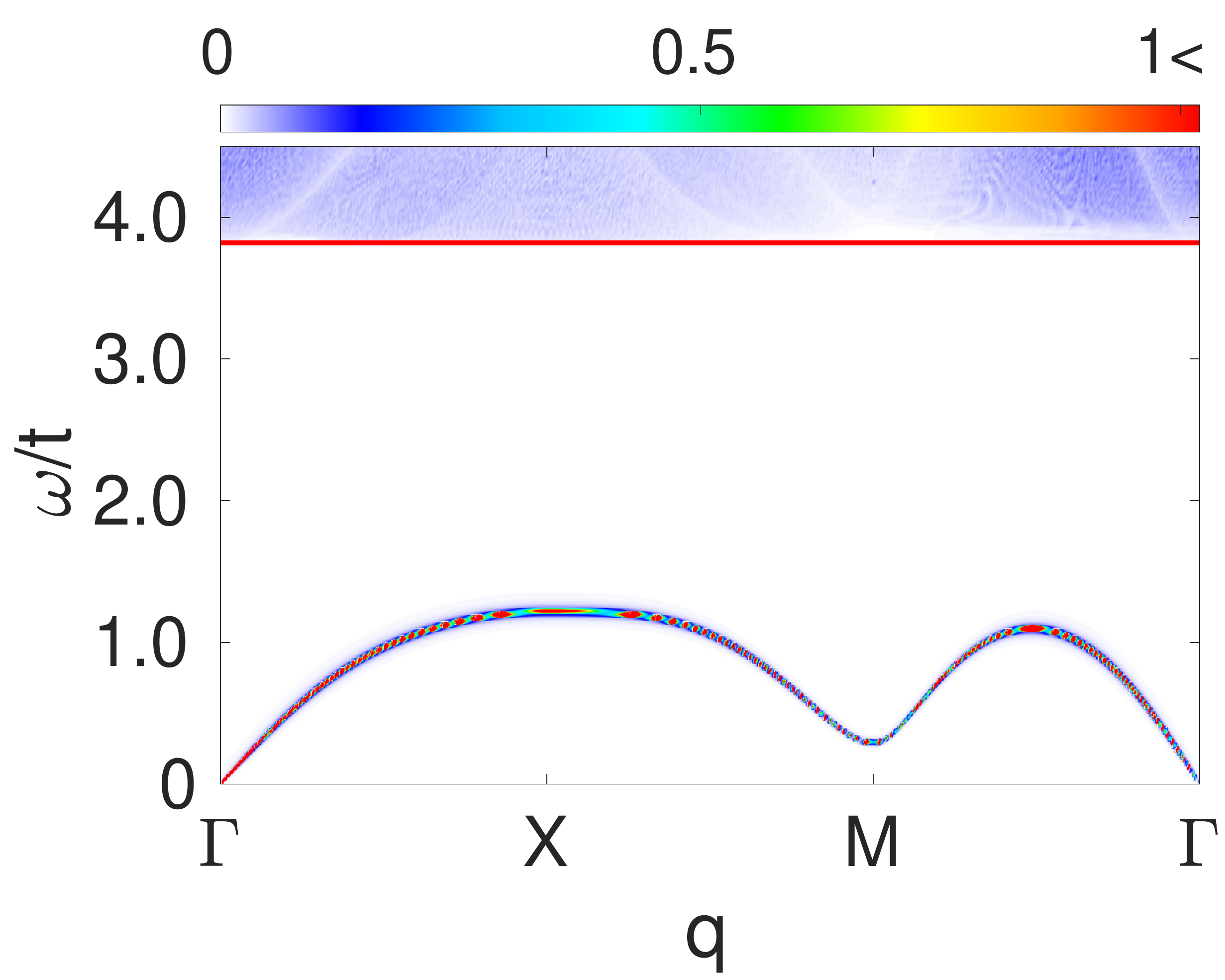}}\\
		\subfigure[Amplitude]{\label{fig.amp_U2}
			\includegraphics[width=6.cm]{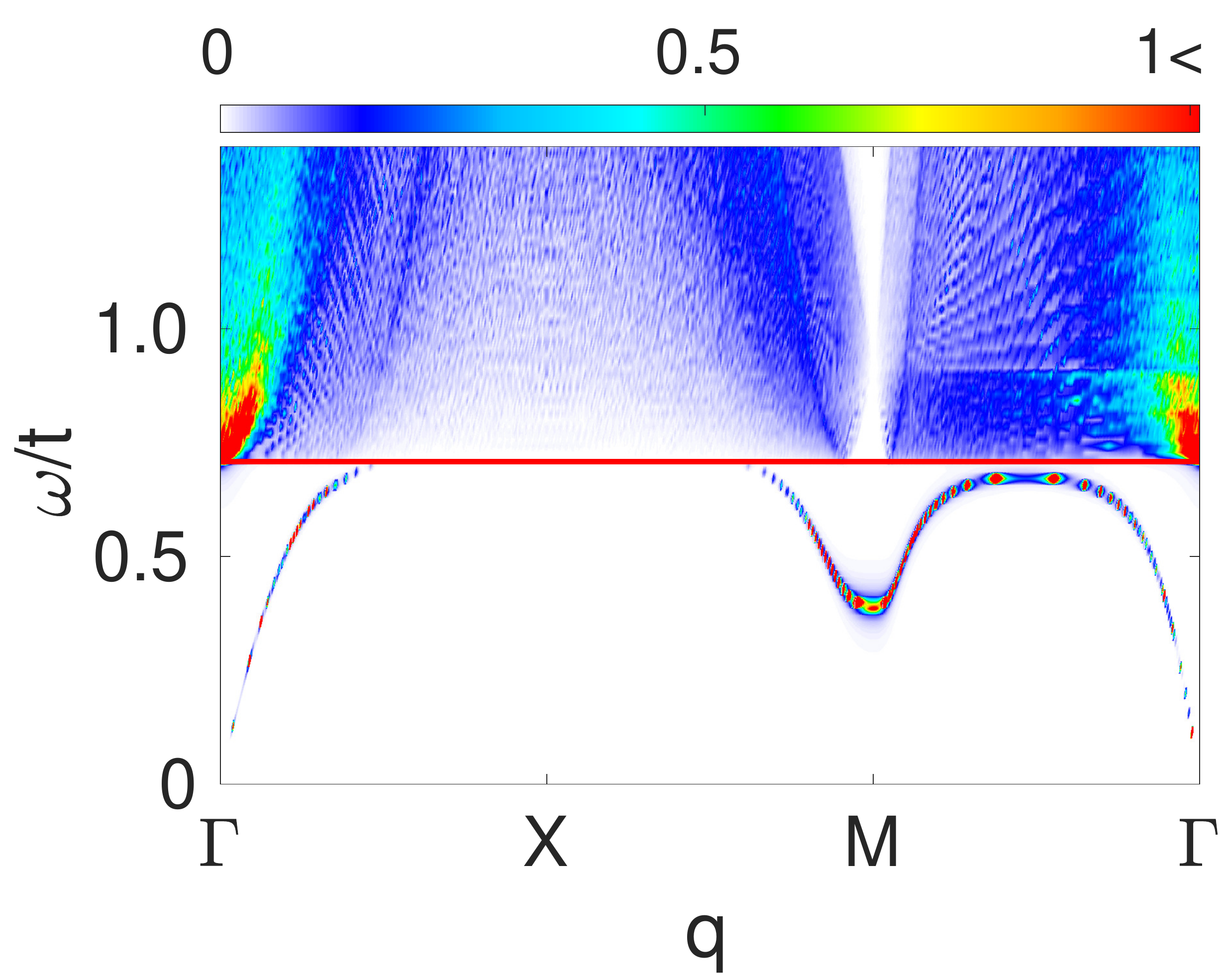}}
		\subfigure[Phase]{\label{fig.pha_U2}
			\includegraphics[width=6.cm]{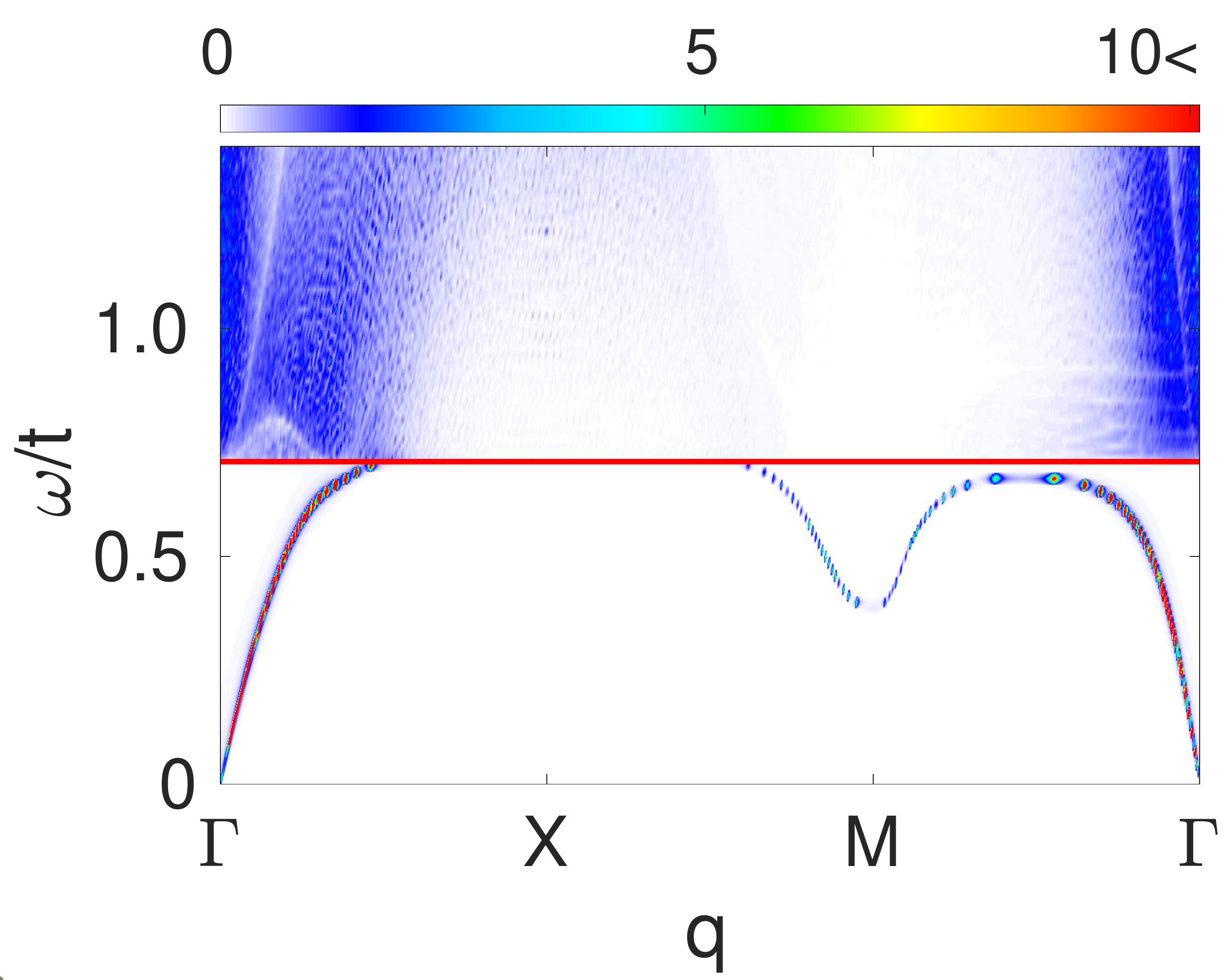}}
		\caption{Amplitude ($P^A(q,\omega)$) and phase ($P^\Phi(q,\omega)$) spectral functions for two different couplings $U$ in the clean limit i.e. $V= 0$, as a function of momenta $q$ (i.e. $\Gamma (0,0)$, $X (\pi,0)$, and $M (\pi,\pi)$ in the momentum space). The figures show the low-energy dispersing collective modes, as well as the two-particle continuum above the spectral gap $\omega_g$ (the red horizontal line). The upper panel is for $U=5$, and the lower panel is for $U=2$. The system size is $200\times200$ and the average density is $\langle n \rangle =0.875$.} \label{Fig.amp_pha_U5_2}
	\end{center}
\end{figure}

\section{II.\,\,\,\,\,\,      Collective modes in clean and disordered cases}
To study the collective modes, we construct the following matrix in real space
\begin{align}
\tilde{\chi}^B = \left(\mathbb{I}_{3N\times 3N}-\chi^B \mathbb{V}\right)^{-1}\chi^B = 
\left(\begin{array}{ccc}
\tilde{\chi}^{AA} & \hat{\chi}^{A\Phi} & \tilde{\chi}^{A\delta n} \\
\tilde{\chi}^{\Phi A} & \tilde{\chi}^{\Phi \Phi} & \tilde{\chi}^{\Phi \delta n} \\
\tilde{\chi}^{\delta n A} & \tilde{\chi}^{\delta n\Phi} & \tilde{\chi}^{\delta n\delta n}
\end{array}\right)
\label{eq.RPA_SM}
\end{align}
to obtain the amplitude spectral function $P_{ij}^A(\omega) = - \frac{1}{\pi}\tilde{\chi}_{ij}^{AA}(\omega) $ and phase spectral function $P_{ij}^\Phi(\omega) = - \frac{1}{\pi}\tilde{\chi}^{\Phi \Phi}_{ij}(\omega) $. Then, we do Fourier transformation to get the $P^A(q,\omega)$ and $P^\Phi(q,\omega)$, to observe the behaviour of amplitude and phase collective modes in momentum space.

In the clean limit, we clearly see the dispersing collective modes in the phase sector, which corresponds to the gapless Goldstone mode, see Fig.~\ref{Fig.amp_pha_U5_2}. However, unlike the Goldstone mode, collective modes in the amplitude sector (the Higgs mode) have a finite gap $\omega_{H}$ \cite{abhisek2020}. In the $q\rightarrow 0$ limit, the gap $\omega_{H}$ is same with the two-particle gap $\omega_g$. In presence of strong coupling, the collective modes are well below the two-particle gap, however as the coupling $|U|$ decreases, the two-particle gap also comes down making the collective modes difficult to observe. The same conclusion also holds as the average density $\langle n\rangle$ decreases.

\begin{figure}[H]
	\begin{center}
		\subfigure[Amplitude]{\label{fig.amp_V0p25}
			\includegraphics[width=6.cm]{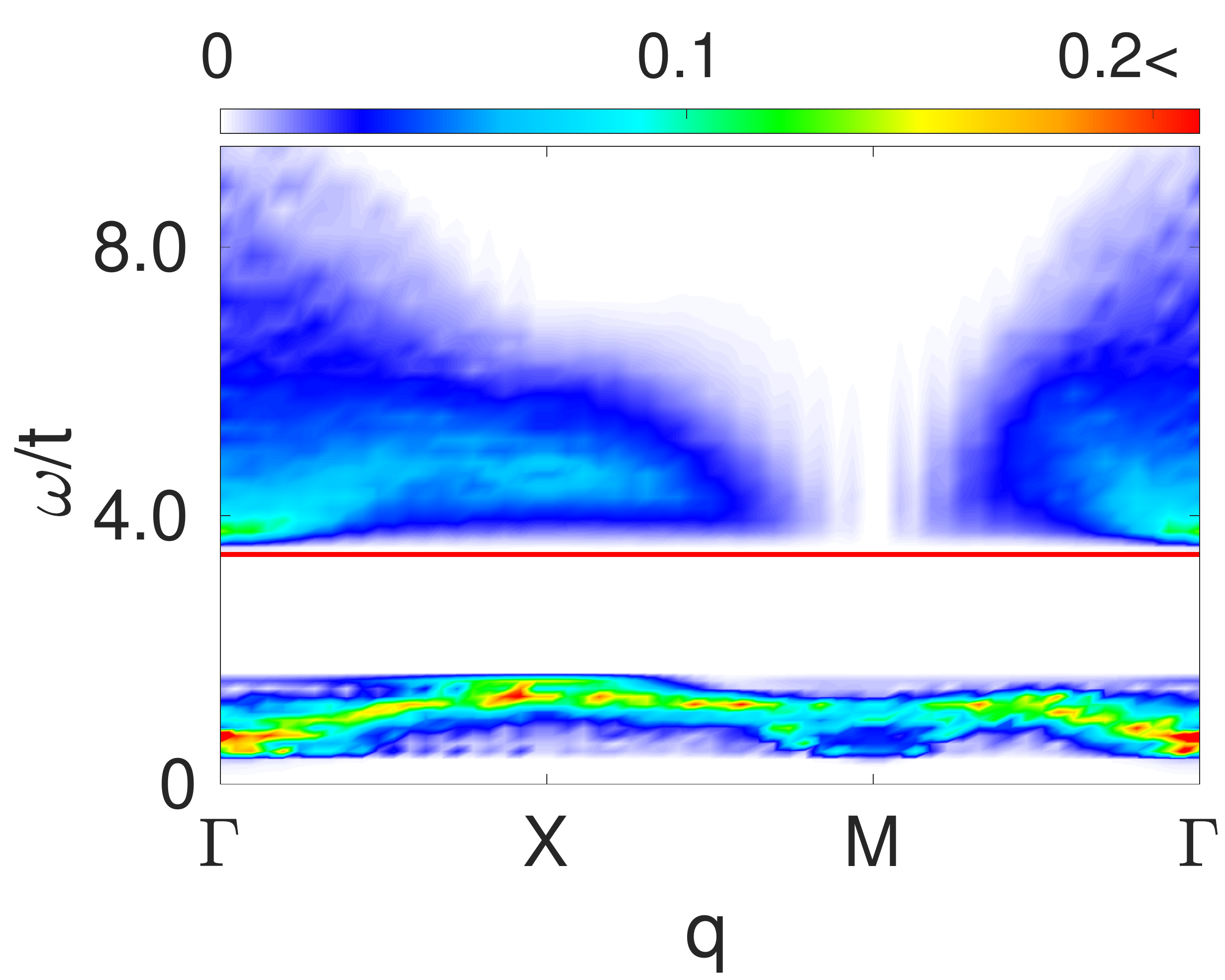}}
		\subfigure[Phase]{\label{fig.pha_V0p25}
			\includegraphics[width=6.cm]{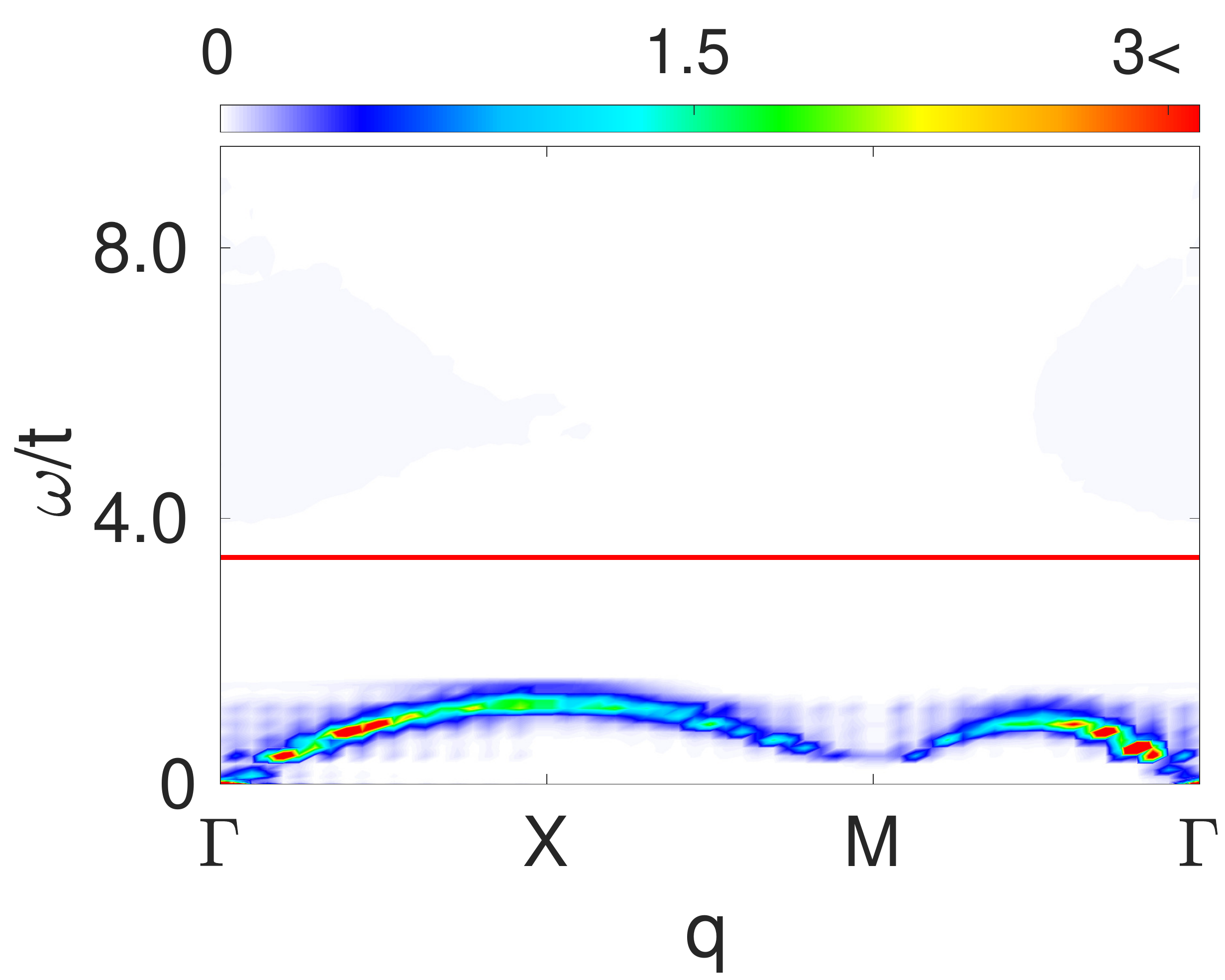}}\\
		\subfigure[Amplitude]{\label{fig.amp_V1p5}
			\includegraphics[width=6.cm]{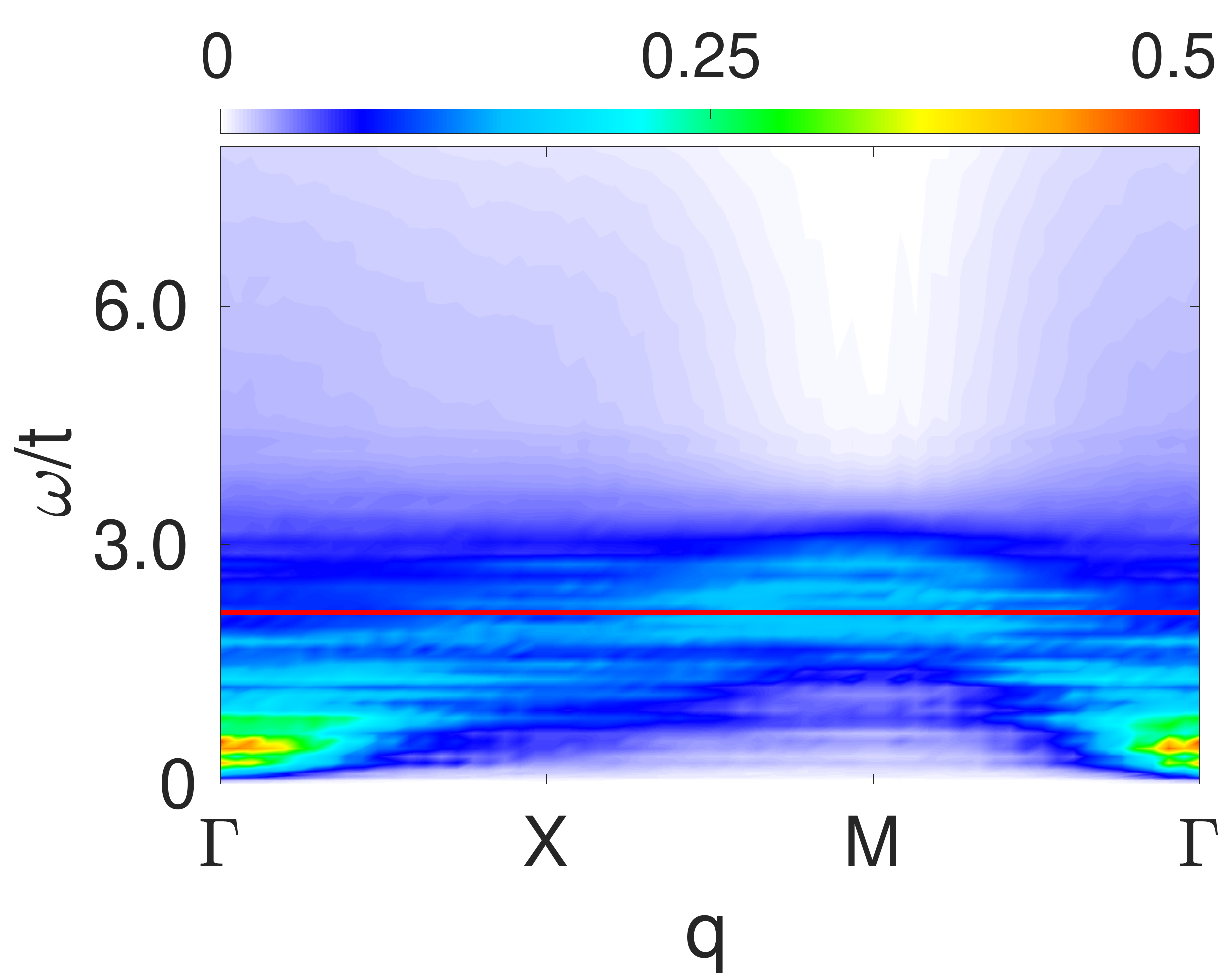}}
		\subfigure[Phase]{\label{fig.pha_V1p5}
			\includegraphics[width=6.cm]{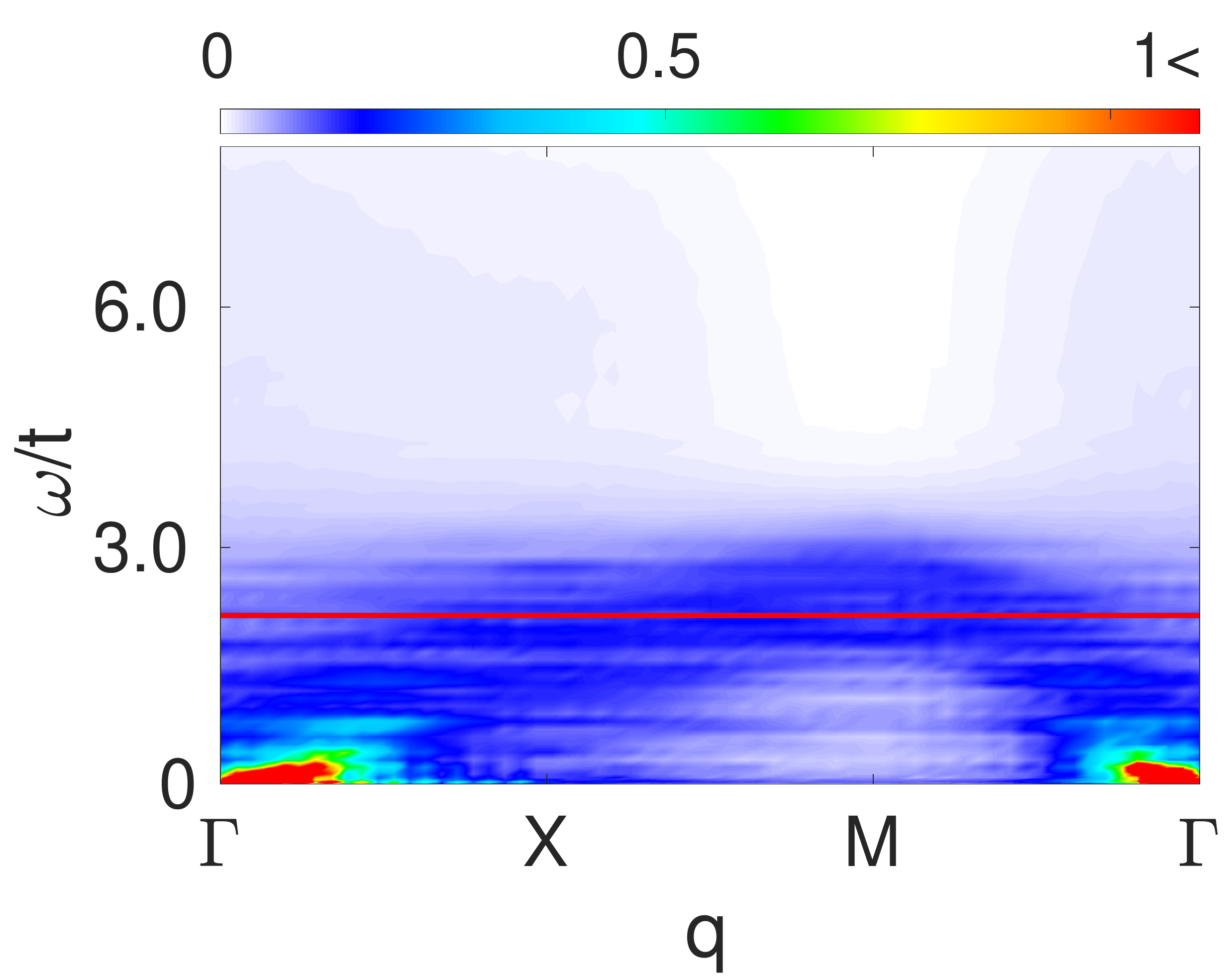}}
		\caption{ Amplitude ($P^A(q,\omega)$) and phase ($P^\Phi(q,\omega)$) spectral functions for a constant coupling $U=5$ in presence of two disorder values, as a function of momenta $q$. The upper panel is for the weak disorder $V=0.25$, and the lower panel is for $V=1.5$. The red horizontal line corresponds to the two-particle spectral gap $\omega_g.$ The system size is $20\times20 $ and the average density is $\langle n \rangle =0.875$.
		} \label{Fig.amp_pha_U5}
	\end{center}
\end{figure}

In presence of weak disorder, in addition to the dispersing collective modes, a non-dispersive mode appears in the amplitude sector at finite energy below two-particle continuum, which has been identified as the \textit{disorder-induced Higgs mode}~\cite{abhisek2020}, see Fig.~\ref{Fig.amp_pha_U5}. On the other hand, the phase mode remains dispersing, but gets broadened. In the strong coupling limit, these modifications are easier to observe as the collective mode structure remains well below the two-particle continuum. When disorder becomes large, the \textit{disorder-induced Higgs mode} gets broadened in energy and hence gets mixed with the incoherent spectral weight coming down from two-particle continuum, and therefore it can not be separately identified. The Goldstone mode is comparatively robust with disorder. However, in presence of sufficiently strong disorder, the sound velocity (related to the slope of the linearly dispersing Goldstone mode) decreases with disorder, and hence the Goldstone mode also ceases to show up as a sharp mode.  

\section{III.\,\,\,\,\,\,      Fluctuation correlation functions and optical conductivity}
\begin{figure}[H]
	\begin{center}
		\subfigure[~Amplitude]{\label{fig.amp_corU5}
			\includegraphics[width=5.5cm]{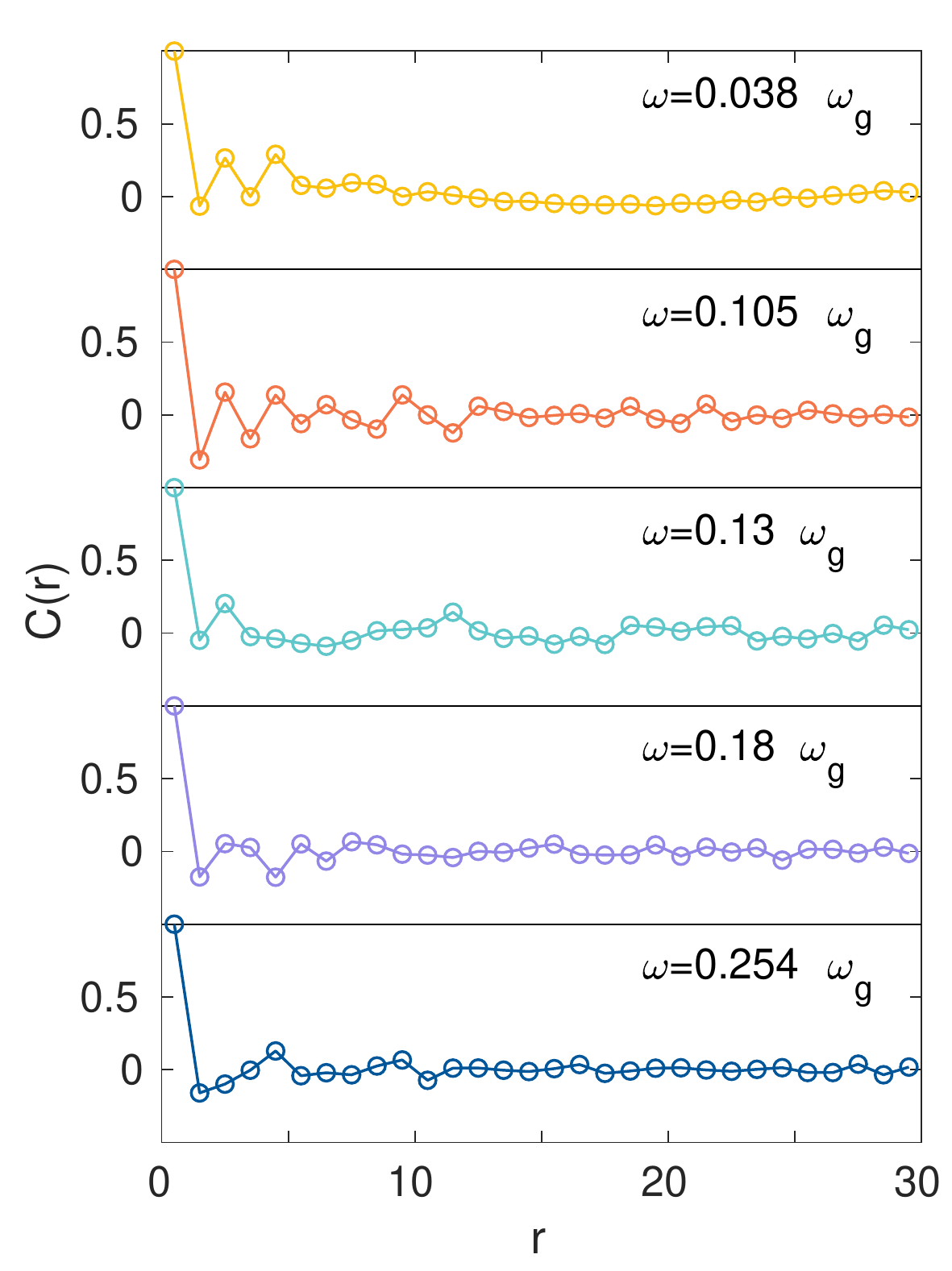}}
		\subfigure[~Phase]{\label{fig.pha_corU5}
			\includegraphics[width=5.5cm]{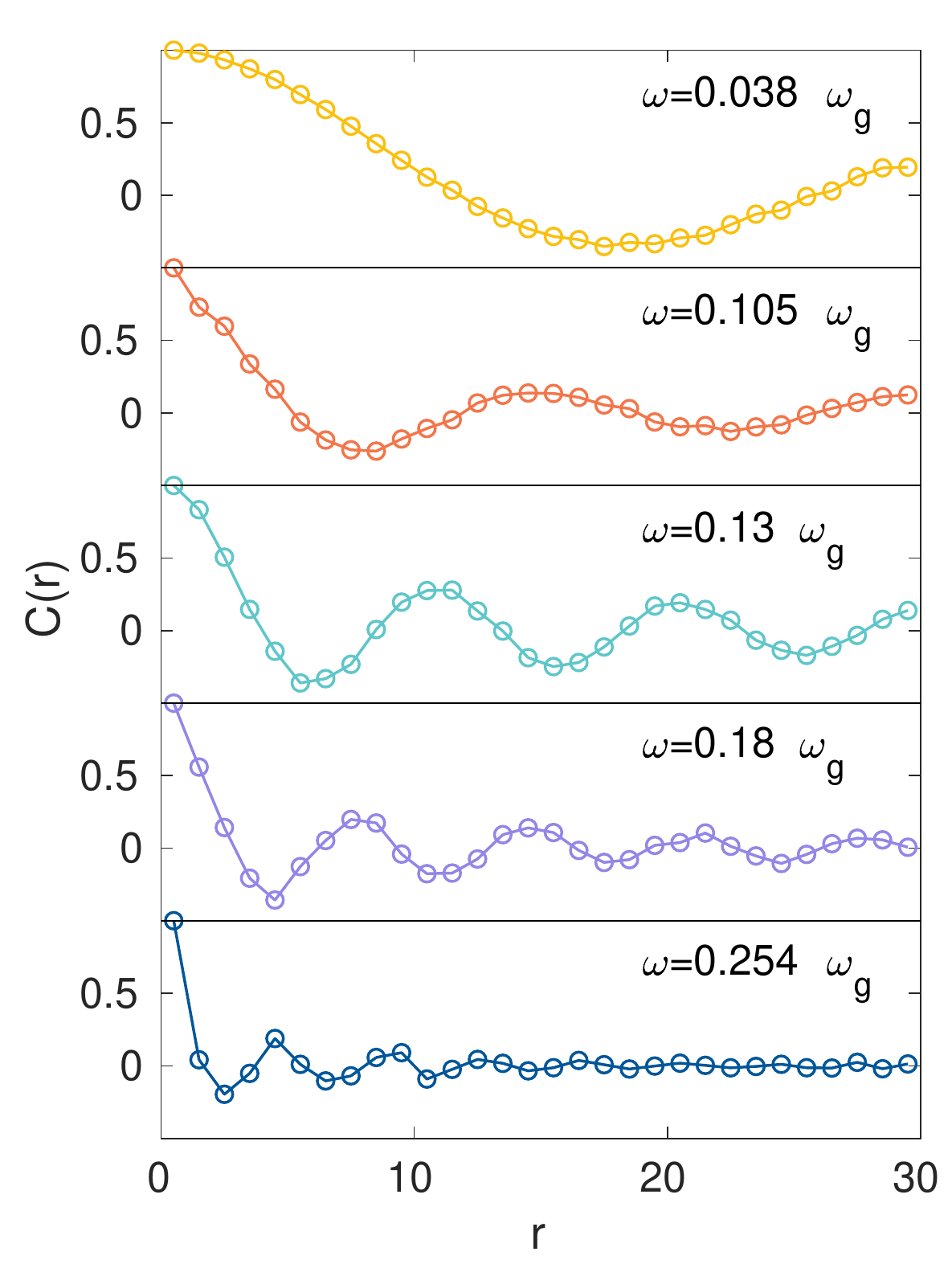}}
		\subfigure[~Phase]{\label{fig.pha_corU2}
			\includegraphics[width=5.5cm]{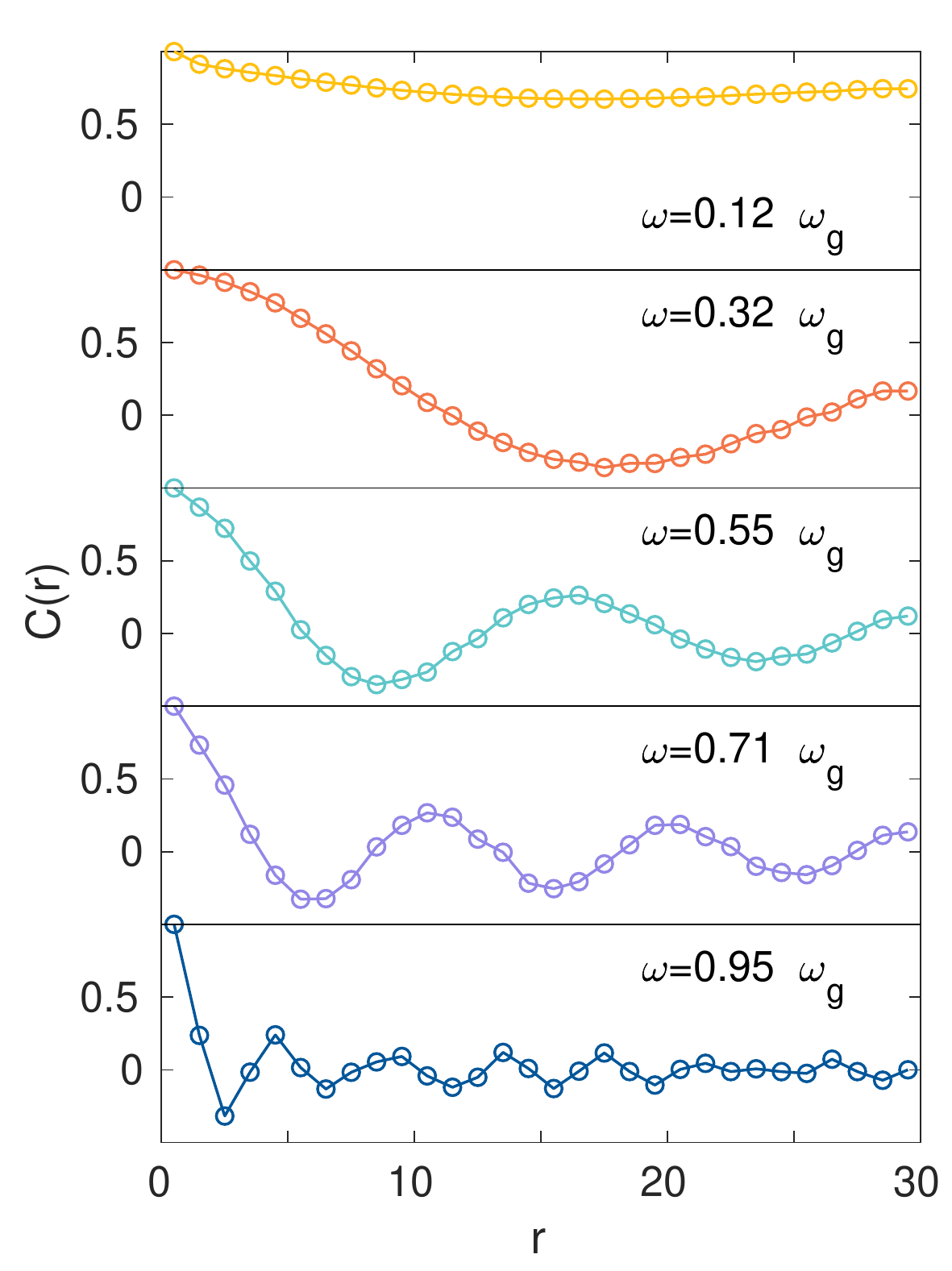}}
		\caption{ \subref{fig.amp_corU5}. The amplitude fluctuation correlation function in presence of a strong coupling $U=5$. \subref{fig.pha_corU5} and \subref{fig.pha_corU2} are the phase fluctuation correlations. \subref{fig.pha_corU5} is for the same strong coupling $U=5$, while \subref{fig.pha_corU2} is for weak coupling $U=2$. The other parameters are $V=0, L=30$, and $\langle n \rangle = 0.875$.} \label{Fig.pha_cor}
	\end{center}
\end{figure}

We now study the amplitude correlation function, which is defined as $C(r)=\langle\tilde{\chi}^{AA}(r,\omega)\rangle/\langle\tilde{\chi}^{AA}(0,\omega)\rangle$. This is shown in Fig.~{\ref{Fig.pha_cor}\subref{fig.amp_corU5}} for the clean system. {The correlation function of amplitude fluctuations decays to $0$ monotonously with a rather short typical length which suggests that its role in the conductivity is limited.}

\begin{figure}[H]
	\begin{center}
		\subfigure[]{\label{fig.cond_U5}
			\includegraphics[width=7.cm]{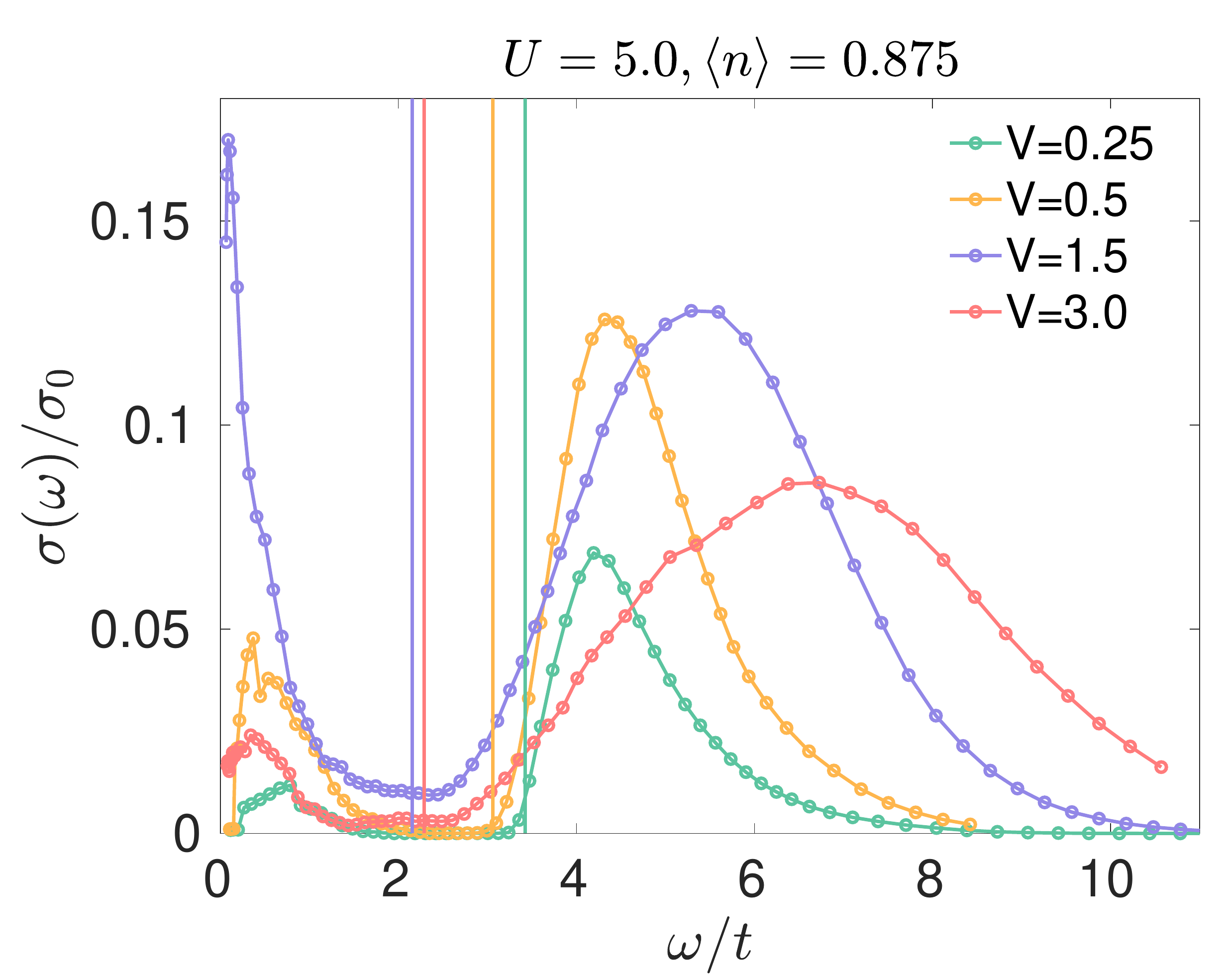}}
		\subfigure[]{\label{fig.cond_U2}
			\includegraphics[width=7.cm]{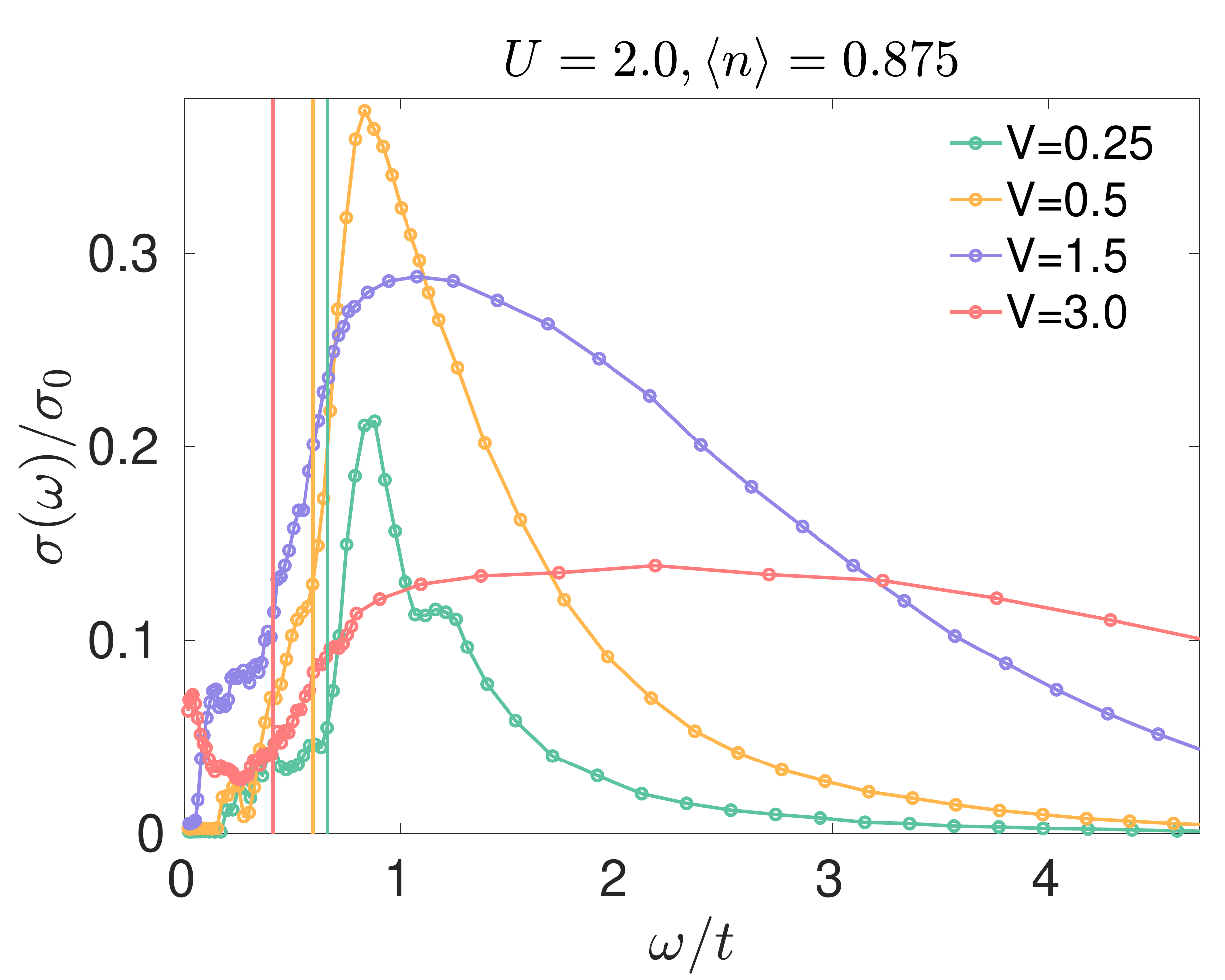}}
		\caption{The optical conductivity $\sigma(\omega)$ in units of $\sigma_0=\frac{e^2}{\hbar}$. The vertical coloured lines are the corresponding two-particle gaps $\omega_g$. } \label{Fig.cond_U5_2}
	\end{center}
\end{figure}

Next we define the phase fluctuation correlation function, defined as $C(r)=\langle\tilde{\chi}^{\Phi\Phi}(r,\omega)\rangle/\langle\tilde{\chi}^{\Phi\Phi}(0,\omega)\rangle$. The phase fluctuation correlation function is rather interesting, which shows a damped oscillation around $C(r)=0$ with increasing distance. We present them in Fig.~{\ref{Fig.pha_cor}\subref{fig.pha_corU5}} and~{\ref{Fig.pha_cor}\subref{fig.pha_corU2}} for two different couplings $U$, where the excitation energy is always below the two-particle spectral gap. {In the strong coupling limit $U=5$, phase fluctuations are excited for even very small energies $\omega \sim 0.04\omega_g$ with a rich oscillating pattern for a broad range of subgap energies. In Fig.~{\ref{Fig.cond_U5_2}\subref{fig.cond_U5}}, the optical conductivity peak is around $\omega \sim 0.18\omega_g$ with some broadening when $V=0.25$. 
For $U=2$ and $\langle n \rangle = 0.875$, the phase fluctuation excites around $\omega \sim 0.32\omega_g$, which is moving to the two particle spectral gap $\omega_g$. The optical conductivity peak is also around $0.6\omega_g$ at weak disorder regime in Fig.~{\ref{Fig.cond_U5_2}\subref{fig.cond_U2}}. }

\end{document}